\documentclass{aa}  

\usepackage{graphicx}
\usepackage{amsfonts}
\usepackage{txfonts}
\usepackage{natbib}
\bibpunct{(}{)}{;}{a}{}{,} 
\begin{document}

\title{High-energy signatures of binary systems of supermassive black holes}

\author{G. E. Romero,{\inst{1}$^,$ \inst{2}} G. S. Vila,\inst{1} \and D. P\'erez\inst{1}   }
		\institute{Instituto Argentino de Radioastronom\'ia, C.C. 5, (1894) Villa Elisa, Buenos Aires, Argentina\\
              \email{danielaperez@iar.unlp.edu.ar, gvila@iar.unlp.edu.ar, romero@iar.unlp.edu.ar}
    \and
    Facultad de Ciencias Astron\'omicas y Geof{\'i}sicas, UNLP, Paseo del Bosque s/n,
    (1900) La Plata, Buenos Aires, Argentina\\
    \email{romero@fcaglp.unlp.edu.ar}}
		
\date{Received ; accepted }

 
  \abstract
   {Binary systems of supermassive black holes are expected to be strong sources of long gravitational waves prior to merging. These systems are good candidates to be observed with forthcoming space-borne detectors. Only a few of these systems, however, have been firmly identified to date. }
   {We aim at providing a criterion for the identification of some supermassive black hole binaries based on the characteristics of the high-energy emission of a putative relativistic jet launched from the most massive of the two black holes. }
   {We study supermassive black hole binaries where the less massive black hole has carved an annular gap in the circumbinary disk, but nevertheless there is a steady mass flow across its orbit. Such a perturbed disk is hotter and more luminous than a standard thin disk in some regions. Assuming that the jet contains relativistic electrons, we calculate its broadband spectral energy distribution focusing on the inverse Compton up-scattering of the disk photons.  We also compute the opacity to the  gamma rays produced in the jet by photon annihilation with the disk radiation and take into account the effects of the anisotropy of the target photon field as seen from the jet.} 
   {We find that the excess of low-energy photons radiated by the perturbed  disk causes an increment in the external Compton emission from the jet in the X-ray band, and a deep absorption feature at energies of tens of TeVs for some sets of parameters.  According to our results, observations with Cherenkov telescopes might help in the identification of supermassive black hole binaries, especially those black hole binaries that host primaries from tens to hundreds of million of solar masses.}
   {}

               
\titlerunning{Binary supermassive black holes at high energies}
\authorrunning{Romero et al.}
               
\maketitle

%

\section{Introduction: Binary black holes}

Binary systems of supermassive black holes are likely the result of mergers of galaxies. The formation and evolution of such systems have been widely discussed in the literature. A recent review of the subject is  given by \citet{Dotti-etal2012}; see also \citet{ColpiDotti2009} and \citet{Komossa2006}. The identification of supermassive black hole binaries (SMBHBs) is currently of great interest, mainly because during the late stages of their evolution these systems are expected to be strong sources of gravitational waves at low and very low frequencies, and thus are suitable targets for space interferometers and pulsar timing arrays \citep{Sesana2013}.

Currently, there are less than 20 identified SMBHBs and a well-populated list of candidates. The known SMBHBs have orbital separations from about $10$ kpc down to $\sim 7$ pc. Those with the largest separations may be directly resolved in X-rays as two distinct active nuclei in the same galaxy \citep{Komossa-etal2003, Fabbiano-etal2011}. Further evidence for SMBHBs includes observations of double broad or narrow lines in quasars \citep{Zhou-etal2004, BorosonLauer2009, Tsalmantza-etal2011, Woo-etal2014} and periodic optical light curves: the BL Lac object \mbox{OJ 287} is the strongest candidate found so far on this basis \citep{Sillanpaa-etal1988, Sillanpaa-etal1996, Valtaoja-etal2000}. Helical distortions, bending, and precession of jets have also been associated with the presence of SMBHBs \citep{Begelman-etal1980, KaastraRoos1992, Roos-etal1993, VillataRaitieri1999, Katz1997, Romero-etal2000, Romero-etal2003, Stirling-etal2003, LobanovRoland2005, LiuChen2007, Caproni-etal2013, Kun-etal2014}. However, binarity is not the only possible explanation of these features and hence many identifications are inconclusive.  The main rival explanation is jet precession by Lense-Thirring effect. In this regard, the search for characteristic features of binarity in the electromagnetic spectrum of SMBHBs, especially during the pre-coalescence phase, has been addressed in a number of studies. These works focus on predicting the line emission and/or the continuum radiation from the accretion flow (e.g. \citealt{Bogdanovic-etal2008, Bode-etal2010, ShenLoeb2010, Zanotti-etal2010, Tanaka-etal2012}). 
 
Active galactic nuclei (AGN) that host SMBHBs at their cores may be in a configuration such that the orbital plane of the binary is coplanar with a circumbinary disk. The coupled evolution of the binary and disk  has been studied by many authors both theoretically and numerically; see for example \citet{GouldRix2000}, \citet{Dotti-etal2007}, \citet{Hayasaki-etal2007}, \citet{Kocsis-etal2012a}, \citet{Roedig-etal2012}, \citet{Dorazio-etal2013}, \citet{Rafikov2013}, and \citet{Hayasaki-etal2012}.  Recent general relativistic magnetohydrodynamic simulations of magnetised circumbinary accretion disks around binary black holes have been performed by \citet{Gold-etal2014a} and \citet{Gold-etal2014b}. The behaviour of a SMBHB embedded in an accretion disk bears some resemblance to a proto-planetary system. In particular, the tidal torques exerted on the gas by the less massive object (the \emph{secondary}) may open a gap (i.e. a region of very low mass density) in the disk (e.g. \citealt{PapaloizouLin1984, LinPapaloizou1986a, LinPapaloizou1986b, Syer-etal1991, SyerClarke1995,Takeuchi-etal1996}).

If the secondary is much less massive than the primary, the disk is hardly perturbed and the inwards migration of the secondary is fast relative to the typical lifetime of the disk. This is called Type I migration. As the mass ratio approaches unity the tidal torques near the secondary are sufficiently strong for the disk to be truncated: an outer ring of gas remains but the inflow of matter across the secondary's orbit is insignificant. For intermediate mass ratios, the inner disk is not evacuated but an annular gap develops about the orbital path of the secondary; this is the Type II migration that proceeds over long timescales.\footnote{The transition between migration regimes depends not only on the mass ratio but also on the orbital separation and properties of the accretion disk such as the value of the viscosity.} The secondary follows the motion of the gas always remaining inside the gap. 

Numerical simulations by \cite{Kocsis-etal2012a, Kocsis-etal2012b} show that, for certain values of the mass ratio and orbital separation, there exists another migration regime characterised by a migration rate that is intermediate between Type I and Type II.  In this so-called Type 1.5 overflowing regime, the gas leaks across the annular gap and the mass density of the disk is different from zero all the way down to the last stable orbit. Then, accretion onto the primary may steadily power an AGN and eventually feed a jet. 

In an accretion disk in the ``overflowing'' regime a considerable amount of gas piles up outside the secondary's orbit, causing the disk to become locally hot and thick.  These perturbations in density and temperature translate into features in the radiative spectrum of the disk that are absent in the case of a standard geometrically thin disk without a gap (e.g. \citealt{LiuShapiro2010, Kocsis-etal2012a, GultekinMiller2012}).

The emission at high energies (i.e. at gamma rays $E_\gamma>1$ MeV) from AGN is radiation entirely produced in the relativistic jets. One of the most efficient mechanisms of high-energy non-thermal emission in jets is the inverse Compton (IC) scattering of photons off relativistic electrons. Low-energy radiation fields emitted both inside and outside the jets are suitable targets for IC interactions; these same photons may absorb the gamma rays produced in the jets. Plenty of target photons are provided by the accretion disk. Since the characteristics of the IC spectrum and the opacity to gamma-ray propagation depend on the energy distribution of the target photons, any noticeable feature in the radiative spectrum of the disk is expected to have a correlation in the high-energy spectrum of the jets. 

In this work, we consider a SMBHB in the ``overflowing'' regime where the primary launches a relativistic jet. We apply the results of the simulations in \citet{Kocsis-etal2012a, Kocsis-etal2012b} to characterise the structure of the disk and calculate its modified radiative spectrum.  We then compute the non-thermal spectral energy distribution (SED) of the jet including the external IC scattering of disk photons and correct it for absorption. Finally, we assess the issue of whether the high-energy part of the SED displays any features that could disclose the presence of a secondary black hole.

The article is organised as follows. In Sect. \ref{sec:disk_model} we provide a brief description of the properties of a relativistic accretion disk with a gap. In Sect. \ref{sec:jet_model} we review the jet model. The spectral energy distributions of the disk and jet for some representative sets of parameters are presented and discussed in Sect. \ref{sec:results}. In Sect. \ref{sect:discussion} we discuss the robustness of our results and their applicability to the identification of SMBHBs. The last brief section is devoted to the conclusions. 

\section{Relativistic accretion disk with a gap}
\label{sec:disk_model}

We consider a SMBHB in steady state. The primary is a Kerr black hole of mass $M_{\bullet}$ and angular momentum $a$ surrounded by an accretion disk. The position of the primary corresponds to cylindrical radius $r=0$. The secondary SMBH of mass $m_{\rm s}$ follows a circular orbit of radius $r_{\rm s}$ immersed in the disk. We assume that the secondary has cleared an annular gap about its orbital path.  In order for this to occur, two basic conditions must be fulfilled: 

\begin{enumerate}

\item The radius of the Hill sphere of the secondary (its region of gravitational influence) must be larger than the disk scale height $h$.\footnote{We assume that the ratio of the disk scale height to its radius, $h/r$, is approximately constant.} The Hill radius is defined as

\begin{equation}
r_{\rm H} \equiv r_{\rm s}\, {\left(\frac{m_{\rm s}}{3M_{\bullet}}\right)}^{\frac{1}{3}} = r_{\rm s}\, {\left(\frac{q}{3}\right)}^{\frac{1}{3}},
\label{eq:hill_radius}
\end{equation}  

where $q = m_{\rm s}/M_{\bullet}$ is the mass ratio. The secondary can open a gap in the disk if $r_{\rm H} \geq h$, which implies from Eq. \ref{eq:hill_radius} that  

\begin{equation}
q\geq 3\left(\frac{h}{r_{\rm s}}\right)^{3}.
\label{eq:gap_opening_condition_1}
\end{equation}  

\vspace{0.2cm}

\item The gap closing timescale due to the viscous reaction of the disk is longer than the gap opening timescale. If the disk is modelled as a standard disk \citep{ShakuraSunyaev1973}, this implies that \citep{Takeuchi-etal1996}

\begin{equation}
q \gtrsim \left(\frac{h}{r}\right)^2 \alpha^{1/2},
\label{eq:gap_opening_condition_2}
\end{equation}  

\noindent where $\alpha$ is the usual dimensionless viscosity parameter.

\end{enumerate}

\begin{figure}[htbp]
\center
\resizebox{\hsize}{!}{\includegraphics{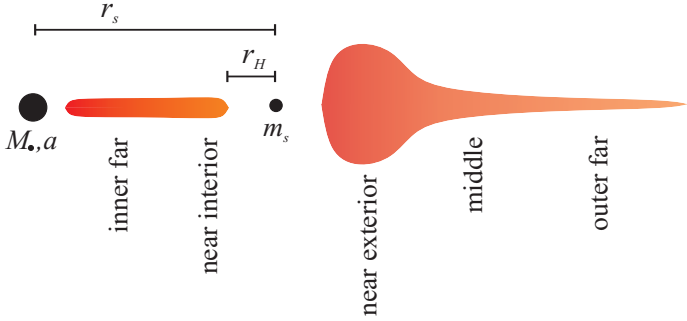}}
\caption{Accretion disk with a gap. The disk is divided into five distinctive zones: the inner and outer far zones, where the perturbations caused by the secondary are negligible, and the middle, near interior, and near exterior zones, where the tidal perturbations of the secondary are strong.}
\label{fig:disk_w_gap}
\end{figure}

For a given value of $M_{\bullet}$, we further restrict the values of $q$ and $r_{\rm s}$ according to the conditions found by \citet{Kocsis-etal2012b} for the system to settle in the ``overflowing'' regime; in particular, we choose $10^{-3} < q < 10^{-1}$ and $r_{\rm s}$ in the range shown in Fig. 3 f of \citet{Kocsis-etal2012a}. The disk may be divided into five distinctive zones as shown in Fig. \ref{fig:disk_w_gap}: inner far, near interior, near exterior, middle, and outer far regions. The influence of the secondary is negligible in the far regions, it is significant in the middle zone, and strong in the near regions. To characterise the structure of the accretion disk in the regions where the tidal perturbations of the secondary are strong, we adopt the prescriptions given by \citet{Kocsis-etal2012b}, which are obtained from fits to the results of their numerical simulations. \citet{Kocsis-etal2012b} compute the surface temperature $T_{\rm s}$ in terms of the central temperature $T_{\rm c}$ and surface density $\Sigma$ of the disk (see Eq. 15 in \citealt{Kocsis-etal2012b}) as
\begin{equation}\label{Ts}
T_{\rm s} = \left(\frac{8}{3 \kappa\Sigma}\right)^{1/4}T_{\rm c},
\end{equation}
where $\kappa = 0.35\;{\rm cm^{2} \: g^{-1}}$ is the opacity assumed to be dominated by electron scattering. The central temperature and surface density are expressed in terms of
\begin{eqnarray}
T_{\rm c} & = & T_{\rm c}\left(\alpha_{-1}, \dot{m}_{-1}, M_{7}, f_{-2}, q_{-3}, r_{{\rm s2}}, r_{2}, r\right),\label{Tc}\\
\Sigma & = & \Sigma\left(\alpha_{-1}, \dot{m}_{-1}, M_{7}, f_{-2}, q_{-3}, r_{{\rm s2}}, r_{2}, r\right),\label{den}
\end{eqnarray}
where:\footnote{\citet{Kocsis-etal2012b} use geometrical units $G = c = 1$.}
\begin{eqnarray}
\alpha_{-1} & = & \frac{\alpha}{0.1},\\
\dot{m}_{-1} & = & \frac{\dot{M}}{0.1 \dot{M}_{\rm Edd}},\\
M_{7}& = & \frac{M_{\bullet}}{10^{7} M_{\odot}},\\
f_{2} & = & \frac{f}{0.01},\\
q_{-3} & = & \frac{q}{10^{-3}},\\
r_{{\rm s2}} & = & \frac{{r_{\rm s}}}{10^{2} M_{\bullet}},\\
r_{2} & = & \frac{r}{10^{2} M_{\bullet}}.
\end{eqnarray}
Here, $\alpha$ is the free parameter in the prescription of \citet{ShakuraSunyaev1973} for the viscosity, $\dot{M}$ is the mass accretion rate, $\dot{M}_{\rm Edd}$ is the Eddington mass accretion rate, and $f$ is a constant calibrated with simulations. In what follows, we describe the main properties of each zone of the perturbed disk.

\begin{itemize}

\item{\textbf{Inner far zone.}} The effects of the secondary SMBH are negligible. The accretion disk, however, is under the strong gravitational field of the primary. We adopt the relativistic disk model of \citet{PageThorne1974} to calculate the temperature profile  in this zone. The expression for the energy flux across the surface of the disk takes the form \citep{NovikovThorne1973, PageThorne1974}

\begin{equation}
F(r) = - \frac{\dot{M}}{4\pi \sqrt{-g}}\,\frac{\Omega,_{r}}{\left(\widetilde{E}-\Omega\widetilde{L}\right)^{2}}
\int^{r}_{r_{{\rm isco}}} \, \left(\widetilde{E}-\Omega\widetilde{L}\right) \widetilde{L},_{r} \,dr,
\label{eq:flux_relativistic_disk}
\end{equation}

where $\Omega$ stands for the angular velocity, and $\widetilde{E}$ and $\widetilde{L}$ represent the specific energy and angular momentum, which can be calculated from the metric coefficients. Assuming the disk radiates as a black body, the surface temperature as a function of the radial coordinate is obtained by means of Stefan-Boltzmann's law,

\begin{equation}
T^{\rm if}_{\rm s}(r) = Z(r)\left(\frac{F(r)}{\sigma_{\rm SB}}\right)^{1/4}.
\label{eq:temp_relativistic_disk}
\end{equation}

Here $\sigma_{\rm SB}$ is the Stefan-Boltzmann constant and $Z$ is a correction by the gravitational redshift (nearly face-on case),

\begin{equation}
Z(r) = \frac{r^{\frac{3}{4}}\left(r^{\frac{3}{2}}+2a\sqrt{r_{\rm g}}-3 r_{\rm g} \sqrt{r}\right)^\frac{1}{2}}{\left(a\sqrt{r_{\rm g}}+r^{\frac{3}{2}}\right)}.
\end{equation}

The gravitational radius of the primary is $r_{\rm g} = GM_{\bullet}/c^{2}$.

\vspace{0.5cm}

\item{\textbf{Near interior zone.}} The near interior zone is immediately inside the orbit of the secondary. Here the tidal effects of the secondary dominate the viscous effects. The surface temperature profile, calculated from Eq. \ref{Ts} and Table 2 from \citet{Kocsis-etal2012b}, is written as

\begin{eqnarray}
T^{\rm ni}_{\rm s} (r)  & = & {\mathbb{C}}^{\rm ni}_{T}\; \frac{8}{3\:\kappa} \; 1.73 \times 10^{4} \; {\dot{m}_{-1}}^{1/4} \; {M_{7}}^{1/4}\; {r_{{\rm s2}}}^{-5/16}\nonumber \\ 
& \times & \left(\frac{r}{10^{2}M_{\bullet}}\right)^{-7/16}{\left|1-\frac{r_{\rm s}}{r}\right|}^{1/4}.
\label{eq:temp_near_interior}
\end{eqnarray}

The value of the constant ${\mathbb{C}}^{\rm ni}_{T}$ is determined by equating Eqs. \ref{eq:temp_relativistic_disk} and \ref{eq:temp_near_interior} at the transition radius $r_{\rm ni}$ between the inner far and near interior zones. We calculate $r_{\rm ni}$ assuming that it is the radial distance where the surface density profiles of the inner far zone  \citep{NovikovThorne1973, PageThorne1974} and near interior zone \citep{Kocsis-etal2012a,Kocsis-etal2012b} take the same value.

\vspace{0.5cm}

\item{\textbf{Near exterior  zone.}} The near exterior zone is immediately outside the orbit of the secondary. As in the near interior zone, tidal effects are more important than viscous effects. From Eq. \ref{Ts} and Table 2 from \citet{Kocsis-etal2012b}, the surface temperature takes the form\footnote{In the near exterior region, we adopt the case of unsaturated tidal torque, that is, $h(r) < r - r_{\rm s}$. In particular, in the approximate expression for the torque per unit mass in the disk given by \citet{Kocsis-etal2012b} in Eq. 5, the function $\Delta$ (Eq. 6) takes the form $\Delta = r -r_{\rm s}$.}

\begin{eqnarray}
T^{\rm ne}_{\rm s} (r) & = & \frac{8}{3 \:\kappa}\;1.12 \times 10^{4}\;{\alpha_{-1}}^{-1/2}\;{f_{-2}}^{5/8}\;{M_{7}}^{-1/8}\;{q_{-3}}^{5/6}\;{r_{\rm s2}}^{19/96}\nonumber \\
& \times & \left(\frac{r}{10^{2}M_{\bullet}}\right)^{-77/96}{\left(\frac{r-r_{\rm s}}{r_{\rm s}}\right)}^{-5/8}{\zeta_{\rm r}}(r)^{1/2},
\label{eq:temp_near_exterior}
\end{eqnarray}

\begin{equation}
{\zeta_{\rm r}} = {\frac{2}{5} \left(\frac{2}{3}\right)}^{\frac{1}{6}} \left(1+\frac{r_{\rm H}}{r_{\rm s}}\right)^{-\frac{115}{72}} 
 \left[1-\left(\frac{r_{\rm s}+r_{\rm H}}{r}\right)^{\frac{115}{72}}\left(\frac{r_{\rm H}}{r-r_{\rm s}}\right)^{\frac{5}{2}}\right].
\end{equation}

Because of intense tidal torques, a significant amount of gas accumulates in the near exterior zone, so the disk is hotter and thicker than a standard thin disk at the same distance from the primary.

\vspace{0.5cm}

\item{\textbf{Middle zone.}} The tidal torque and heating are locally negligible compared to the viscous effects. There is, nonetheless, a considerable amount of gas accumulated. The temperature profile yields (see Eq. \ref{Ts} and Table 2 from \citealt{Kocsis-etal2012b})

\begin{eqnarray}
T^{\rm m}_{\rm s} (r) & = & {\mathbb{C}}^{\rm m}_{T}\;\frac{8}{3\: \kappa}\;2.39 \times 10^{4}\; {\alpha_{-1}}^{-1/2}\;{M_{7}}^{-1/8}\;{f_{-2}}^{5/8}\;{r_{\rm s2}}^{1/16}\nonumber \\
& \times & \left(\frac{r}{10^{2}M_{\bullet}}\right)^{-7/8}.
\label{eq:temp_middle}
\end{eqnarray}

As in the near interior zone, the value of the constant ${\mathbb{C}}^{\rm m}_{T}$ is calculated by equating Eqs. \ref{eq:temp_near_exterior} and \ref{eq:temp_middle} at the transition radius $r_{\rm ne}$ between the near exterior and middle zones. An expression for $r_{\rm ne}$ is given in \citet{Kocsis-etal2012b}.

\vspace{0.5cm}

\item{\textbf{Outer far zone.}} The outermost part of the disk is basically unperturbed by the secondary and may be described by the Shakura-Sunyaev model. The surface temperature has the well-known expression (see Eq. \ref{Ts} and Table 2 from \citealt{Kocsis-etal2012b})

\begin{eqnarray}
T^{\rm of}_{\rm s} (r) & = & \frac{8}{3\:\kappa} \; 2.06 \times 10^{4}\;{\dot{m}_{-1}}^{1/4}\;{M_{7}}^{-1/4}\nonumber \\
& \times & \left(\frac{r}{10^{2}M_{\bullet}}\right)^{-3/4}\;\left[1-\sqrt{\frac{r_{\rm isco}}{r}}\,\right]^{1/4},
\label{eq:temp_outer_far}
\end{eqnarray}

where $r_{\rm isco}$ is the radius of the innermost stable circular orbit. 

\end{itemize}

\section{Jet model}
\label{sec:jet_model}

To characterise the jet, we apply the model developed by \citet{RomeroVila2008}, \citet{Reynoso-etal2011}, and \citet{Vila-etal2012}.  This type of jet model has been extensively applied to galactic microquasars in the low hard state, but has also been adapted to FR I radio galaxies \citep{Reynoso-etal2011} and blazars \citep{Reynoso-etal2012}. Since the jet is parametrised in terms of the accretion rate, scaling for different situations is straightforward.The general prescriptions, such as the evolution of the magnetic field with the distance to the black hole or the existence of a particle acceleration zone in the region where the outflow is plasma dominated, are expected to have a wide range of validity.

\begin{figure}[htbp]
\center
\resizebox{\hsize}{!}{\includegraphics{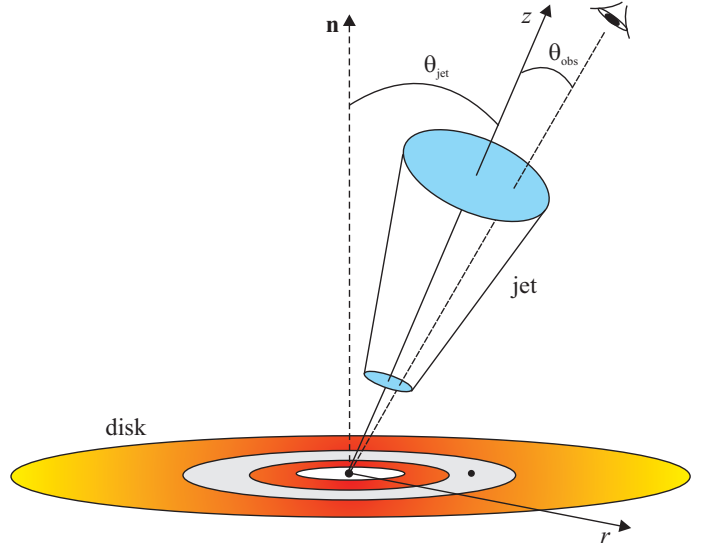}} 
\caption{Sketch of the disk-jet system (not to scale). Some relevant geometrical parameters are shown.}
\label{fig:sketch_jet}
\end{figure}

The jets are assumed to be two conical outflows launched from a distance $z_0 = 50 r_{\rm g}$ from the primary with an initial radius  $r_0 = 0.1 z_0$.  The symmetry axis of the jets, which we define as the $z$-axis, is inclined an angle $\theta_{\rm jet}$ with respect to the normal to the disk, and forms an angle $\theta_{\rm obs}$ with the line of sight of the observer;\footnote{For simplicity, we assume that the normal to the disk, $z-$axis, and line of sight lie in the same plane. This simplification does not have any impact on the results of the calculations.} see Fig. \ref{fig:sketch_jet} for a sketch. We assume that the outflows propagate with a constant bulk Lorentz factor $\Gamma_{\rm jet}$ up to a distance $z_{\rm end}$ from the primary before braking because of the interaction with the external medium.

Each jet carries a total power \mbox{$L_{\rm jet} = 0.05\dot{M}c^2$}. In the region $z_{\rm acc} \leq z \leq z_{\rm max}$ (hereafter the \emph{acceleration region}),  a fraction  \mbox{$L_{\rm rel} = 0.05L_{\rm jet}$} of this power is transferred to particles that are accelerated up to relativistic energies by some mechanism we do not specify. Here we only consider the injection of electrons, although jets containing relativistic protons have been studied in previous applications of the same model; see \citet{Vila-etal2012} and references therein. 

We adopt an injection function of relativistic electrons that, in the reference frame co-moving with the jet,  is isotropic and a power law with an exponential cutoff in energy,

\begin{equation}
Q(E,z) =  Q_0\, E^{-p} \exp\left[-E/E_{\rm max}(z)\right]  \quad [Q] = \rm{erg}^{-1} \rm{s}^{-1} \rm{cm}^{-3}.
\label{eq:injection_rel_particles}
\end{equation}

\noindent The function $Q$ is different from zero only in the acceleration region and for $E\geq E_{\rm min}$. The maximum energy $E_{\rm max}(z)$ is determined by the balance between the total energy loss rate and the acceleration rate. We include adiabatic and radiative energy losses by relativistic Bremsstrahlung, synchrotron radiation, synchrotron-self Compton (SSC), and external inverse Compton scattering (EC) with the accretion disk photons as targets. For the acceleration rate we adopt the usual expression

\begin{equation}
	\left.\frac{dE}{dt}\right|_{\rm{acc}} = \eta\,e\,c\,B(z).
\end{equation}\label{eq:acceleration_rate}

\noindent Here $c$ is the speed of light, $e$ the electron charge, and $\eta <1$ is a dimensionless parameter that characterises the efficiency of the acceleration mechanism. The magnetic field strength in the jet decays as $B(z) \propto z^{-1}$.

Finally, to calculate the distribution of relativistic electrons in steady state in the co-moving reference frame, $N(E,z)$, we numerically solve the kinetic equation

\begin{equation}
v_{\rm{conv}}\frac{\partial N}{\partial z} + \frac{\partial}{\partial E} \left(\left.\frac{dE}{dt}\right|_{\rm{tot}}N\right) = Q(E,z) \quad [N] = \rm{erg}^{-1} \rm{cm}^{-3}
\label{eq:transport_equation}
\end{equation}

\noindent in the region $z_{\rm acc} \leq z \leq z_{\rm end}$. In this equation, $\left.dE/dt\right|_{\rm{tot}}$ is the sum of all energy loss rates, and $v_{\rm{conv}}\sim v_{\rm{jet}}$ is the convection velocity of the plasma. 

 We compute the SED of the jet\footnote{For all radiative processes the SED is initially computed in the co-moving frame and then transformed to the reference frame of the observer, except for external Compton (see below).}  applying the usual formulae for synchrotron radiation, relativistic Bremsstrahlung, and inverse Compton scattering in the Thomson and Klein-Nishina regimes (e.g. \citealt{BlumenthalGould1970, RomeroVila2014}). In the case of the external Compton spectrum, we account for the anisotropy of the radiation field of the disk as seen from the jet; see for instance \citet{AharonianAtoyan1981}, \citet{DermerSchlickeiser1993}, \citet{Dermer-etal2009}, and \citet{Khangulyan-etal2014}. The general expression for the EC emissivity (in units of erg$^{-1}$ s$^{-1}$ sr$^{-1}$) of photons with energy $E_\gamma$ scattered into solid angle $\Omega_{\gamma}$  is given by

\begin{eqnarray}
q_{\rm EC}\left(E_\gamma, \Omega_\gamma, z\right) & = & c \int d\Omega_{\rm ph} \int dE_{\rm ph} \, \frac{dN_{\rm ph}}{dE_{\rm ph}d\Omega_{\rm ph}}\,\times \\ \nonumber && \int d\Omega_e \int dE\, (1-\beta_e\cos\psi) \, \frac{dN}{dE d\Omega_e}\,\frac{d\sigma_{\rm IC}}{dE_\gamma d\Omega_\gamma}.
\label{eq:qEC_complete}
\end{eqnarray}

\noindent Here $dN_{\rm ph}/dE_{\rm ph}d\Omega_{\rm ph}$ is the number density of target (disk) photons with energy $E_{\rm ph}$ propagating in the direction given by the solid angle $\Omega_{\rm ph}$,  $dN/dE d\Omega_e$ is the number density of electrons with energy $E$ propagating in the direction of $\Omega_e$, the factor $\beta_e = \sqrt{1 -\gamma_e^{-2}}\sim 1$ where $\gamma_e$ is the electron Lorentz factor, $\psi$ is the collision angle, and $d\sigma_{\rm IC}/dE_\gamma d\Omega_\gamma$ is the Klein-Nishina double differential cross section.

\noindent  We approximate $\Omega_e \sim \Omega_\gamma$, since $\gamma_e\gg 1$ and thus the photons are basically scattered in the direction of motion of the electrons. Under this assumption the double differential cross section simplifies to

\begin{equation}
\frac{d\sigma_{\rm IC}}{dE_\gamma d\Omega_\gamma} \sim \frac{d\sigma_{\rm IC}}{dE_\gamma}\,\delta\left(\Omega_\gamma - \Omega_e\right),
\label{eq:ic_cross_section_head_on}
\end{equation}

\noindent where $d\sigma_{\rm IC}/dE_\gamma$ is given by Eqs. 25-27 in \citet{Dermer-etal2009} or Eq. 2 in \citet{Khangulyan-etal2014}. Then Eq. (\ref{eq:qEC_complete}) reduces to

\begin{eqnarray}\label{eq:qEC_head_on}
q_{\rm EC}\left(E_\gamma, \Omega_\gamma, z\right) & = & c \int d\Omega_{\rm ph} \int^{E_{\rm ph}^{\rm max}}_{E_{\rm ph}^{\rm min}} dE_{\rm ph} \, \frac{dN_{\rm ph}}{dE_{\rm ph}d\Omega_{\rm ph}}\,\times \\ \nonumber &&  \int dE\, (1-\beta_e\cos\psi) \, \frac{dN}{dE d\Omega_e}\,\frac{d\sigma_{\rm IC}}{dE_\gamma}
\end{eqnarray}

\noindent with the collision angle and the electron distribution evaluated at $\Omega_e=\Omega_\gamma$. The limits of the integral over the target photon energy are

\begin{equation}
E_{\rm ph}^{\rm min} = \left(\frac{E_\gamma}{E-E_\gamma}\right)\frac{m_ec^2}{2\gamma_e(1-\beta_e\cos\psi)} \quad E_{\rm ph}^{\rm max} =\frac{2E_\gamma}{(1-\beta_e\cos\psi)}.
\label{eq:limits_target_photons}
\end{equation}

\noindent The integrals in Eq. \ref{eq:qEC_head_on} may be equivalently computed  in a reference frame fixed to the accretion disk (i.e. the observer frame) or in the jet co-moving frame. We choose to perform the calculations in the observer frame where the disk photon field is known as follows:

\begin{equation}
\left.\frac{dN_{\rm ph}}{dE_{\rm ph}d\Omega_{\rm ph}}\right|_{\rm obs}= \frac{2}{c^3h^3}\frac{E_{\rm ph}^2}{\exp[E_{\rm ph}/(kT_{\rm s})]-1}.
\label{eq:target_photon_distribution}
\end{equation}

\noindent The (isotropic) electron distribution in the jet frame is

\begin{equation}
\left.\frac{dN}{dE d\Omega_e}\right|_{\rm jet} = \frac{N}{4\pi},
\label{eq:electron_distribution}
\end{equation}

\noindent where $N$ is given by the solution of Eq. \ref{eq:transport_equation}. It may be transformed to the observer frame applying Eq. 5  of \citet{TorresReimer2011}.

Finally, to correct for absorption, the SED is multiplied by an attenuation parameter $\exp (-\tau_{\gamma\gamma})$, where $\tau_{\gamma\gamma}(E_\gamma)$ is the optical depth for pair production in two-photon annihilation (e.g. \citealt{BeckerKafatos1995}). This is a first order correction and the development of an electromagnetic cascade remains to be investigated.

\section{Results}
\label{sec:results}

We consider the two sets of parameters given in Table \ref{tab:general_parameters}, $M_{\rm disk1}$ and $M_{\rm disk2}$, to characterise the accreting system. The mass of the primary is $10^7M_{\odot}$ in model $M_{\rm disk1}$, and $10^8M_{\odot}$ in model $M_{\rm disk2}$. In both cases, the angular momentum of the primary is $a/r_{\rm g} = 0.99$, which corresponds to $r_{\rm isco}=1.4545\:r_{\rm g}$, and the mass accretion rate is $\dot{M} = 0.1 {\dot{M}}_{\rm Edd}$.  We notice that microquasars in the low hard state and blazars are usually thought to be inefficient accreting sources. Advection-dominated accretion flow (ADAF) and truncated disk-plus corona models are commonly adopted in the description of these kinds of sources. The accretion rate adopted here is in the upper band of the allowed values, which corresponds to the more luminous sources (see e.g. \citealt{Narayanetal1998}).

The values of $r_{\rm s}$ and $q$ are chosen to obtain the largest possible orbital separation and the widest gap. These parameters are not independent. Figure 4 of \citet{Kocsis-etal2012b} shows the allowed region in the $q-r_{\rm s}$ space. We approximate the half-width of the  gap by the Hill radius $r_{\rm H}$ of the secondary. The simulations of \citet{Kocsis-etal2012b,Kocsis-etal2012a} do not apply inside the gap, so the region \mbox{$r_{\rm s} - r_{\rm H} < r < r_{\rm s} + r_{\rm H}$} is excluded from our calculations. The remaining two free parameters in the formulas of \citet{Kocsis-etal2012b} are the viscosity parameter $\alpha$ and a coefficient $f$ in the expression of the tidal torque. We fix $\alpha = 0.1$ and $f=0.01$ following \citet{Kocsis-etal2012b}.

\subsection{Temperature profile and radiative spectrum of the disk}

\begin{figure*}[htbp]
\center
\includegraphics[width = 0.48\textwidth, keepaspectratio]{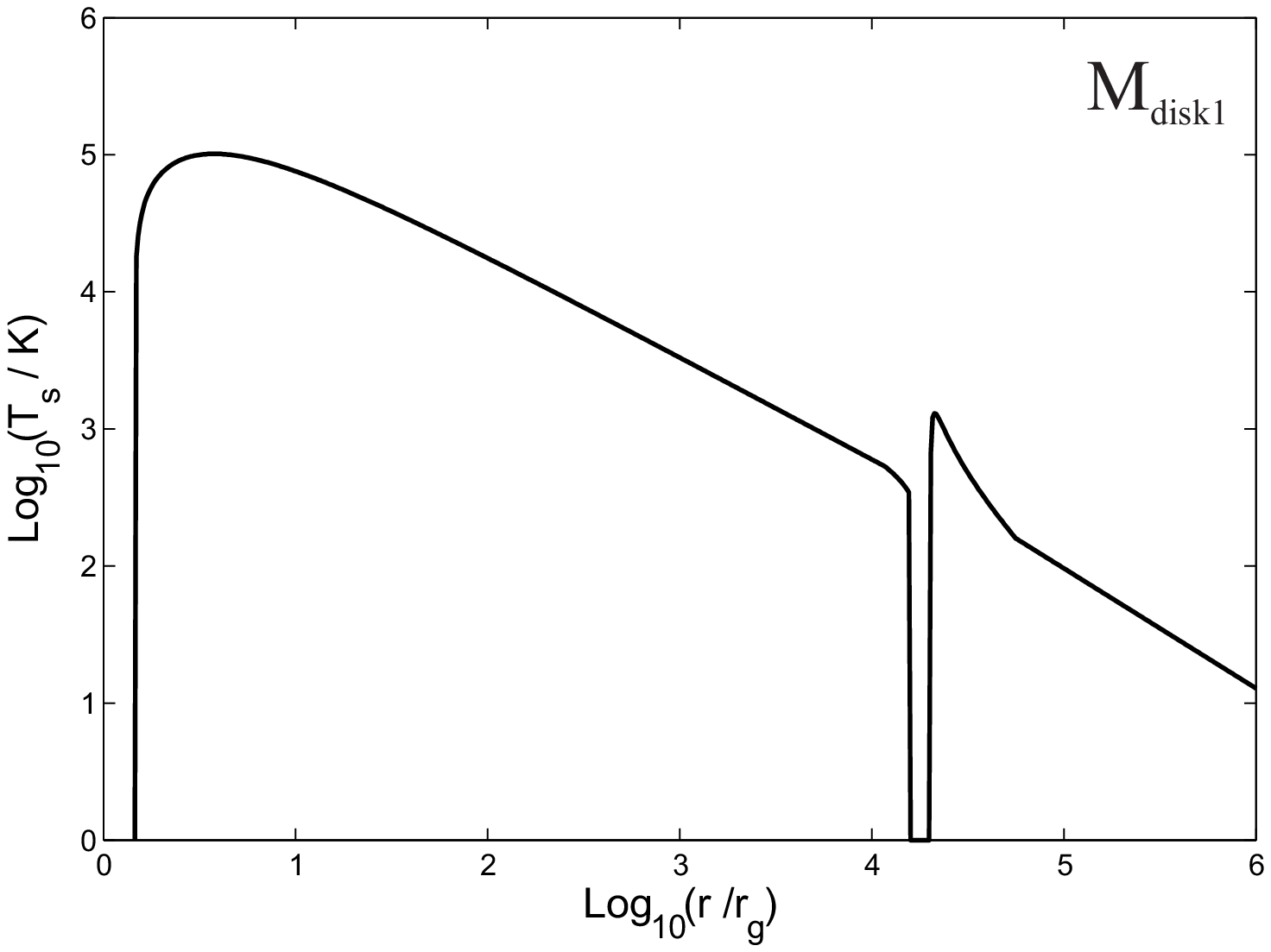}
\includegraphics[width = 0.48\textwidth, keepaspectratio]{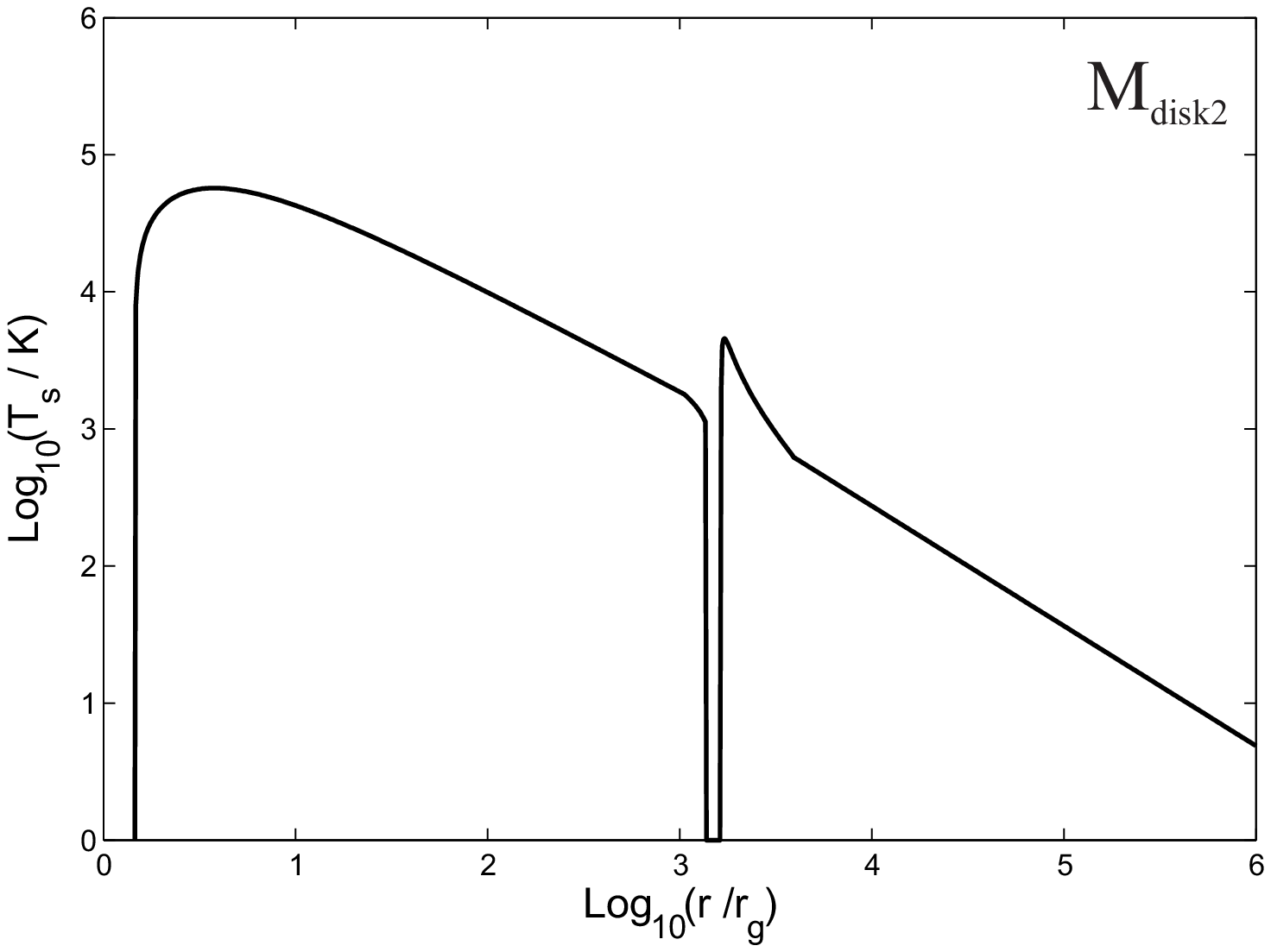}
\caption{Surface temperature of the disk as a function of radius for the set of parameters $M_{\rm disk1}$ (left) and $M_{\rm disk2}$ (right).}
\label{fig:disk_temperature}
\end{figure*}
   
\begin{figure*}[htbp]
\center
\includegraphics[width = 0.48\textwidth, keepaspectratio]{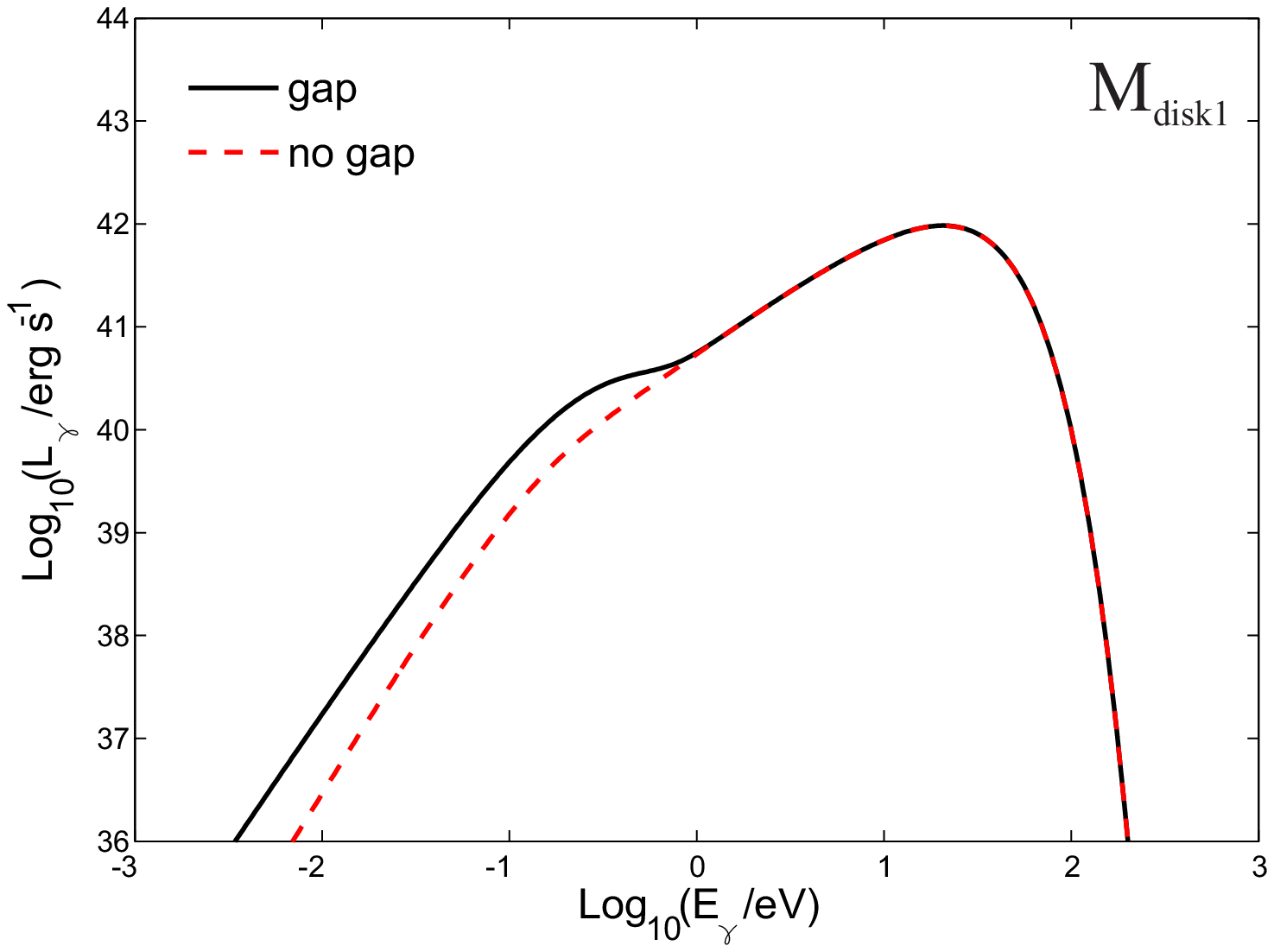}
\includegraphics[width = 0.48\textwidth, keepaspectratio]{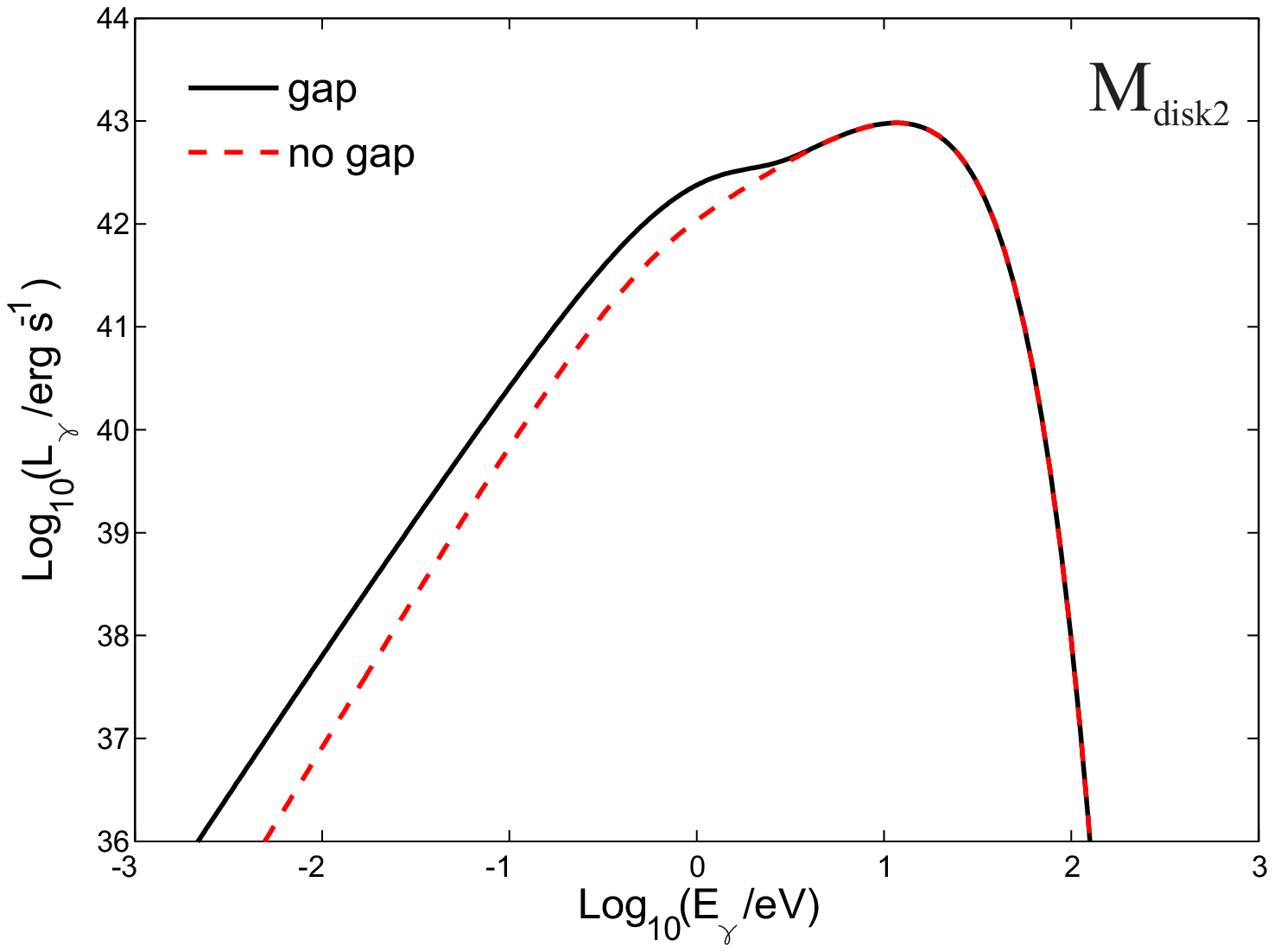}
\caption{Spectral energy distribution of the disk for the set of parameters $M_{\rm disk1}$ and $M_{\rm disk2}$ (right). The luminosity of a relativistic accretion disk without a gap around a SMBH with the same mass, spin, and accretion rate as the primary is shown for comparison in dashed line.  In all SEDs shown in the figures we plot the total luminosity as a function of energy.}
\label{fig:sed_disk}
\end{figure*}

Figure \ref{fig:disk_temperature} shows the surface temperature of the disk as a function of radius for both models, calculated with the parametrisation given in Sect. \ref{sec:disk_model}. Because of the increased density and pressure, the temperature rises sharply just outside the secondary's orbit. The corresponding spectral energy distributions  of the disk are plotted in Fig. \ref{fig:sed_disk}. The spectra are clearly different from those of a standard accretion disk without a gap around a black hole of the same mass and angular momentum as the primary. At low energies, the luminosity is higher than that of an accretion disk without a gap. This increment is caused by the high temperatures in the near exterior and middle zones compared with the temperature in the Page-Thorne model. From about 1 eV (near-infrared) up to the cutoff at $\gtrsim 100$ eV (ultraviolet) both SEDs coincide. As the mass of the primary increases and the distance between the SMBHs decreases, the size of the gap gets smaller and, hence,  the signatures of the gap on the SED begin to disappear.  We also notice that the perturbation caused by the secondary occurs at large distances from the black hole in comparison with the distances associated with the inner disk. Hence, if the disk is truncated and its inner part is replaced by an ADAF, this would not significantly affect the high-energy effects we discuss in this work.

\subsection{Non-thermal radiation from the jet}

\begin{figure*}[htbp]
\center
\includegraphics[width = 0.48\textwidth, keepaspectratio]{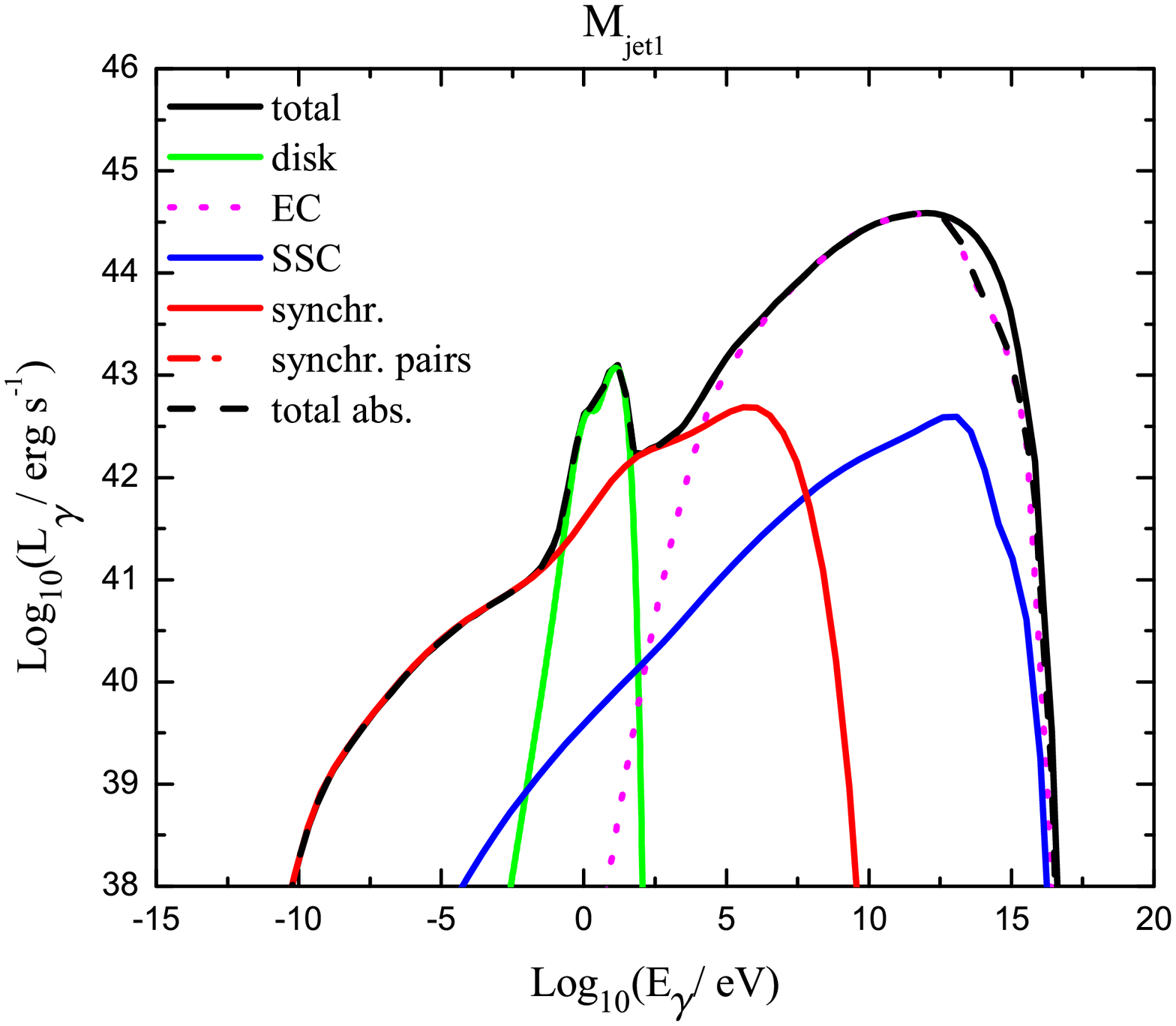}
\includegraphics[width = 0.48\textwidth, keepaspectratio]{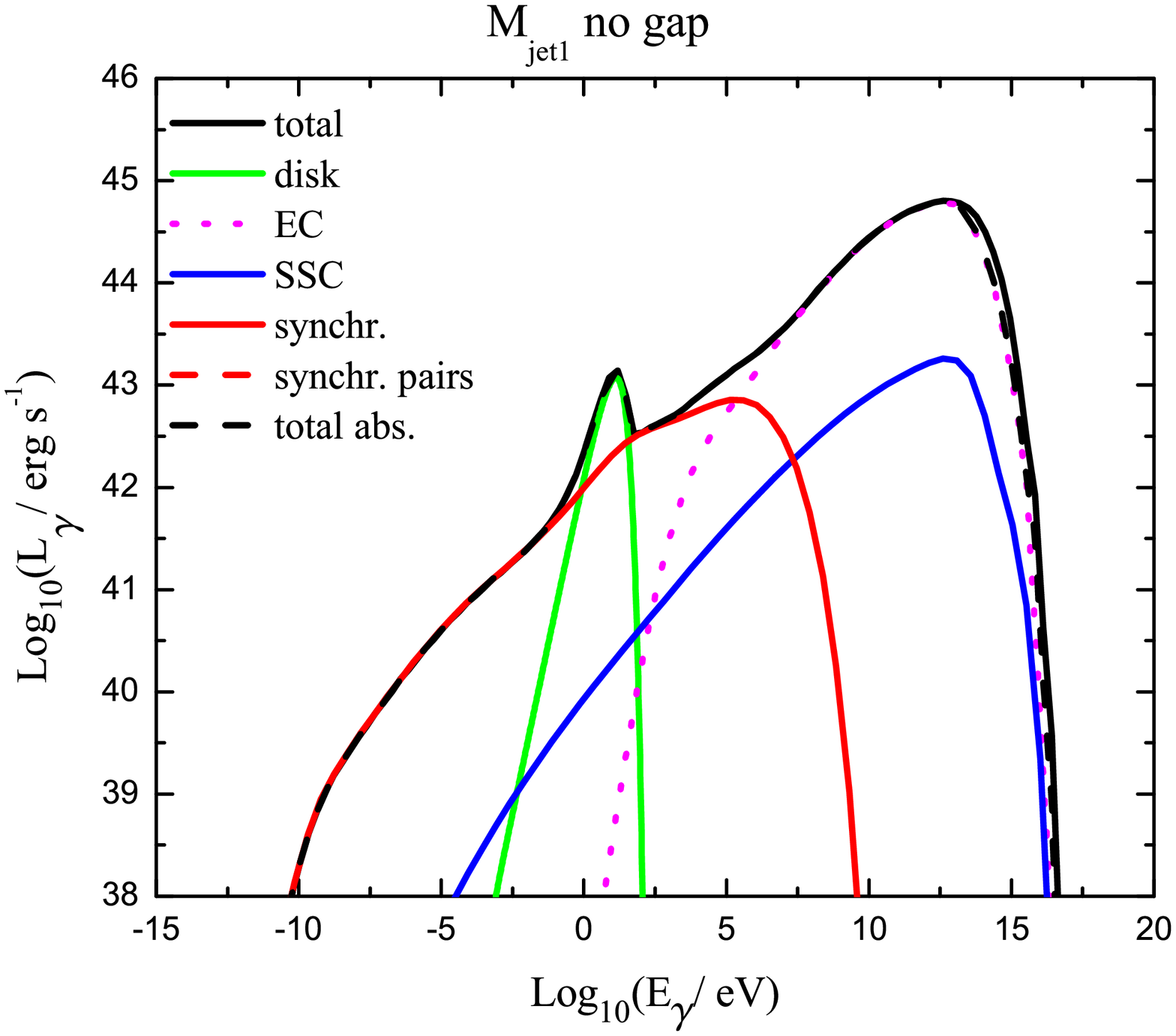}
\caption{Left: spectral energy distribution of the jet corrected for absorption for the set of parameters $M_{\rm jet1}$. Right: the same but for the case of a relativistic thin disk without a gap around a black hole of the same mass, spin, and accretion rate as the primary.  }
\label{fig:sed_jet_M1}
\end{figure*}

\begin{figure*}[htbp]
\center
\includegraphics[width = 0.48\textwidth, keepaspectratio]{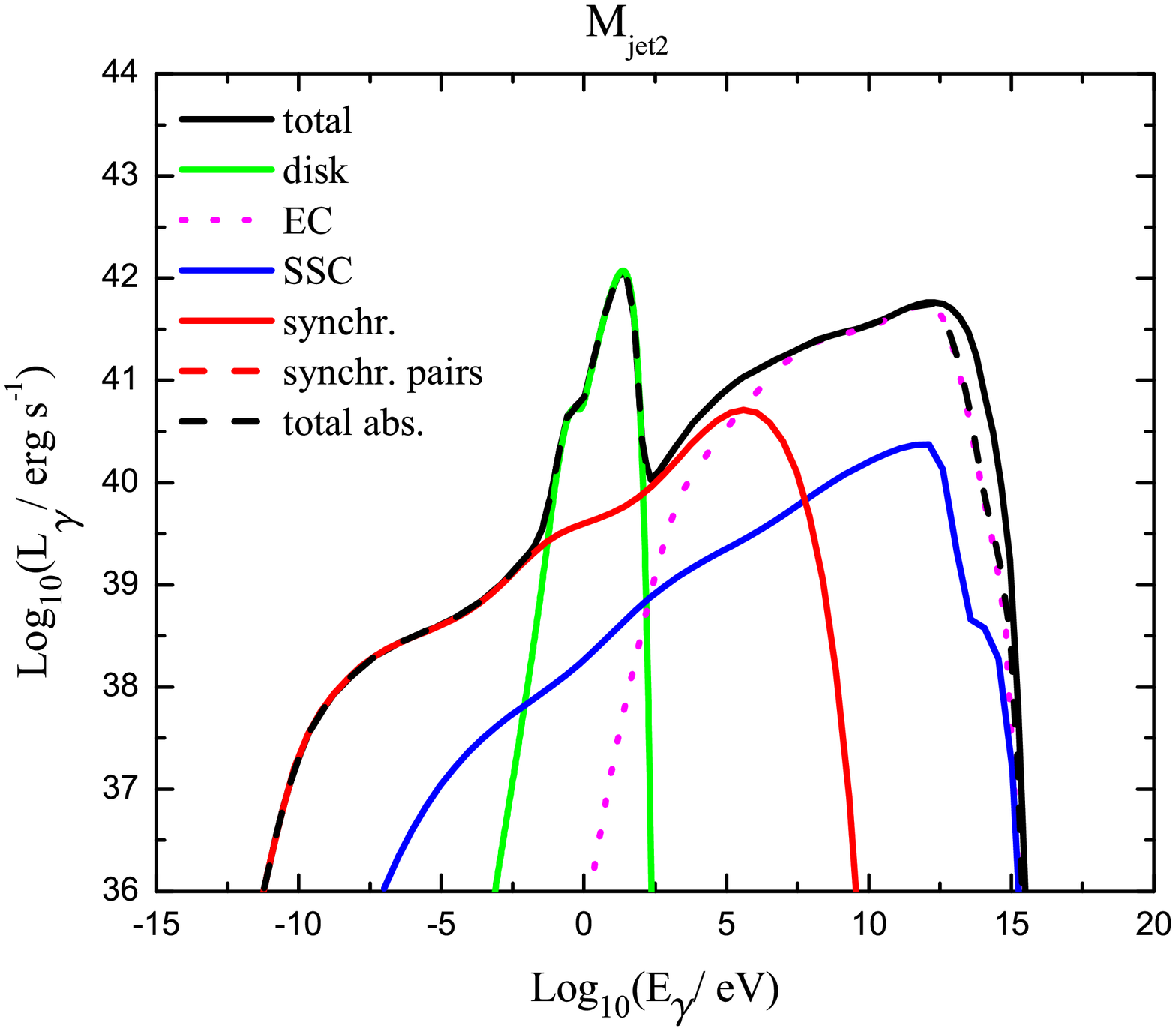}
\includegraphics[width = 0.48\textwidth, keepaspectratio]{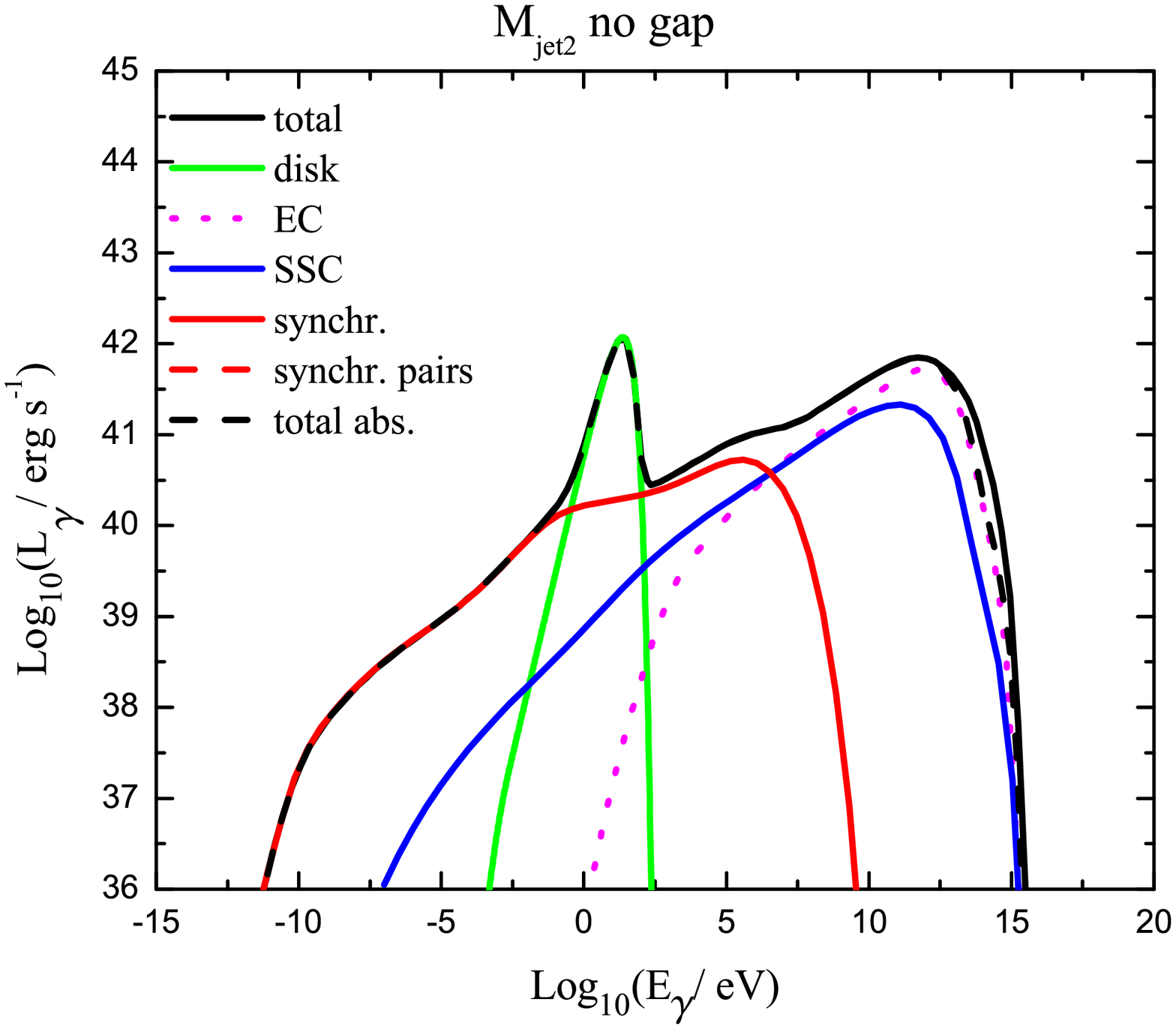}
\caption{Left: spectral energy distribution of the jet corrected for absorption for the set of parameters $M_{\rm jet2}$. Right: the same but for the case of a relativistic thin disk without a gap around a black hole of the same mass, spin, and accretion rate as the primary.  }
\label{fig:sed_jet_M2}
\end{figure*}

\begin{figure*}[htbp]
\center
\includegraphics[width = 0.48\textwidth, keepaspectratio]{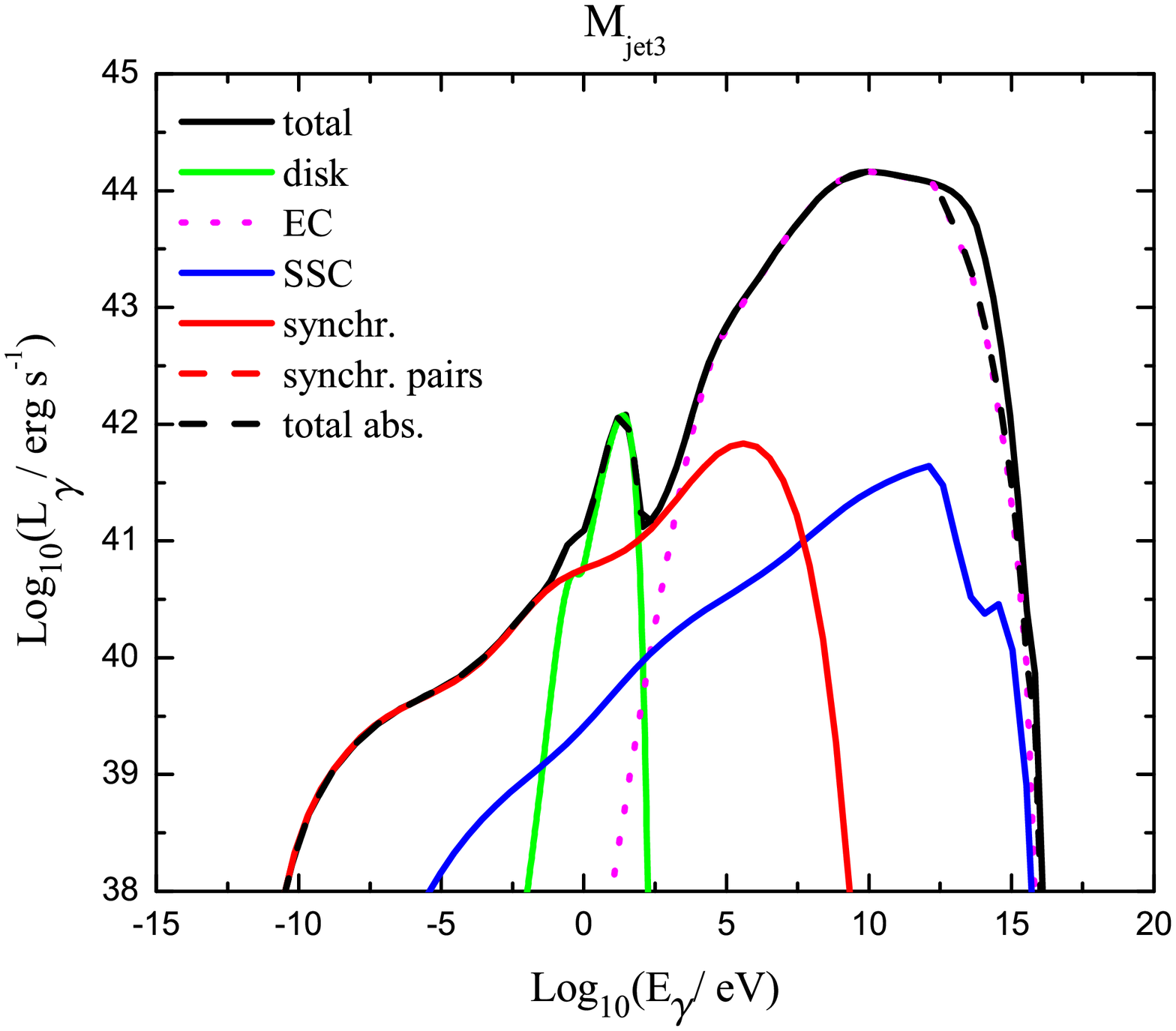}
\includegraphics[width = 0.48\textwidth, keepaspectratio]{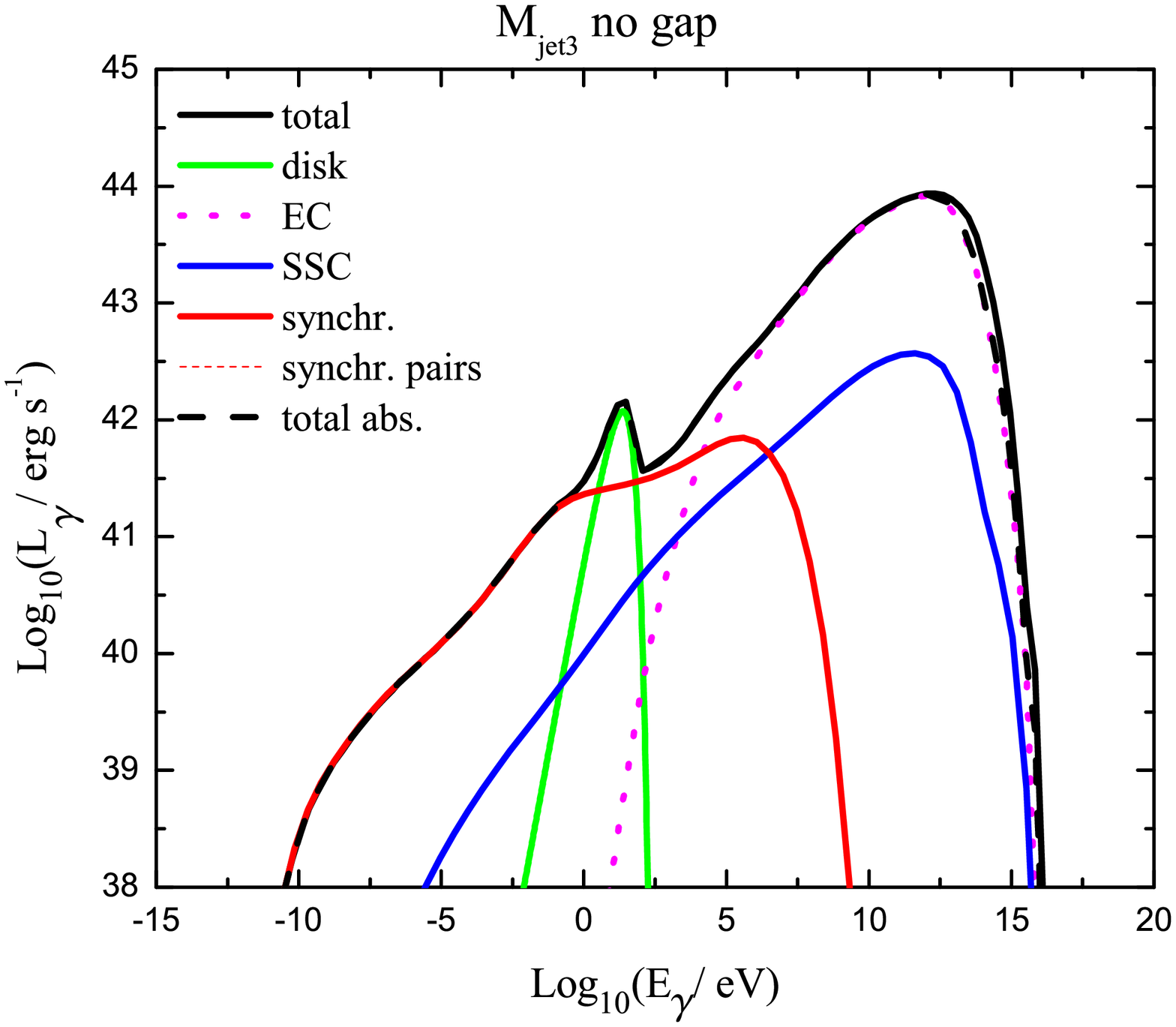}
\caption{Left: spectral energy distribution of the jet corrected for absorption for the set of parameters $M_{\rm jet3}$. Right: the same but for the case of a relativistic thin disk without a gap around a black hole of the same mass, spin, and accretion rate as the primary.  }
\label{fig:sed_jet_M3}
\end{figure*}

\begin{figure*}[htbp]
\center
\includegraphics[width = 0.48\textwidth, keepaspectratio]{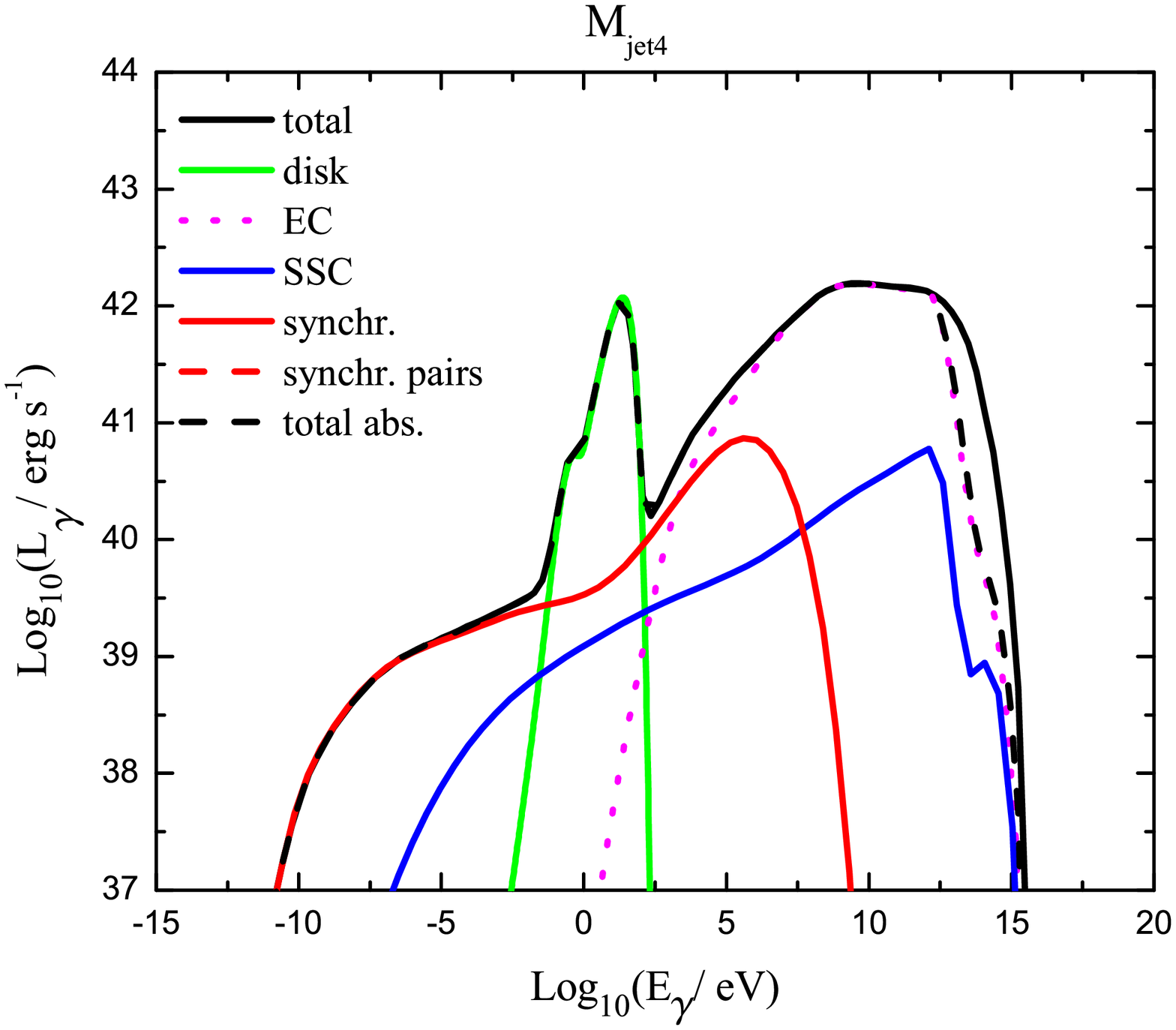}
\includegraphics[width = 0.48\textwidth, keepaspectratio]{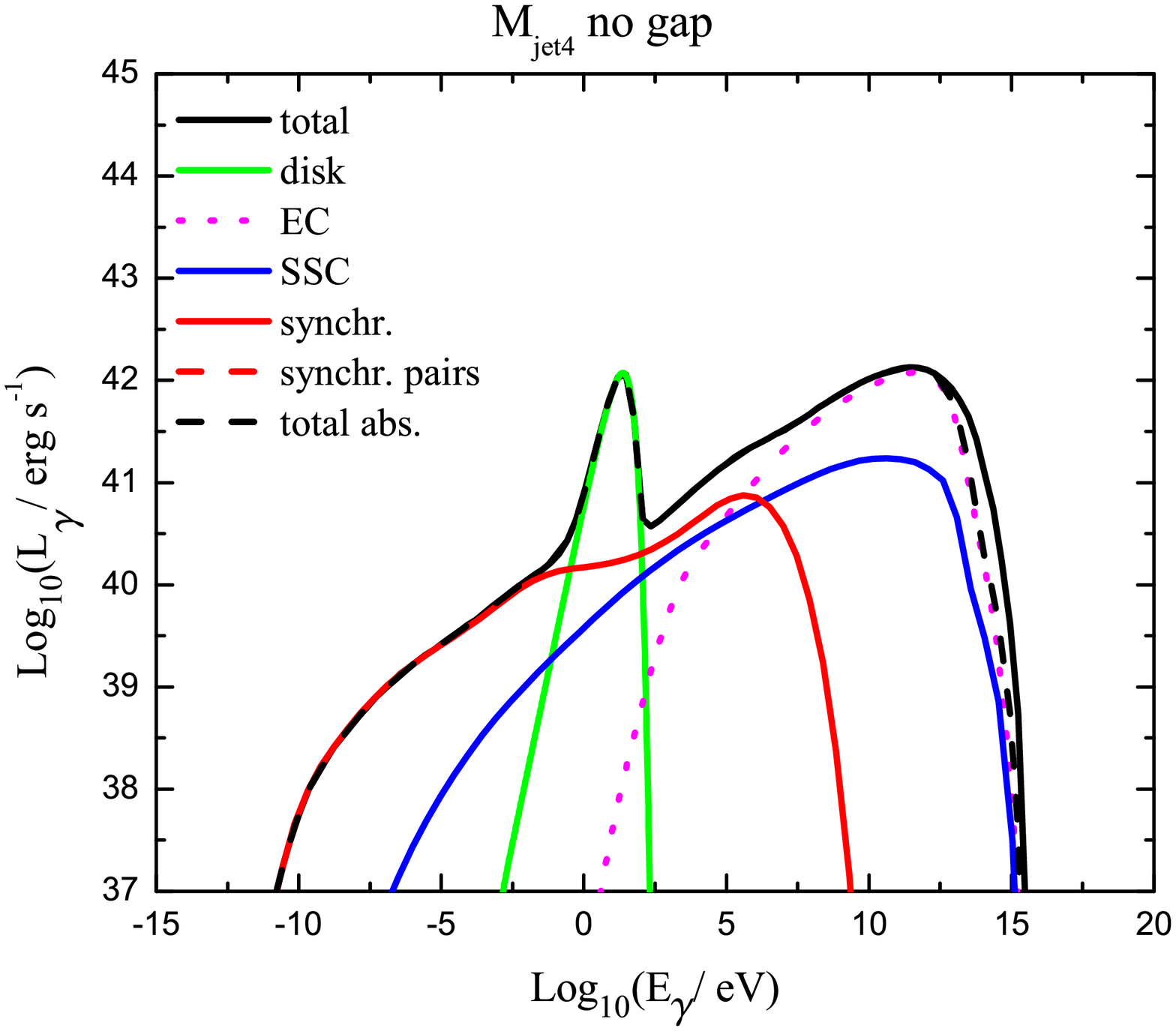}
\caption{Left: spectral energy distribution of the jet corrected for absorption for the set of parameters $M_{\rm jet4}$. Right: the same but for the case of a relativistic thin disk without a gap around a black hole of the same mass, spin, and accretion rate as the primary.  }
\label{fig:sed_jet_M4}
\end{figure*}

\begin{figure*}[htbp]
\center
\includegraphics[width = 0.48\textwidth, keepaspectratio]{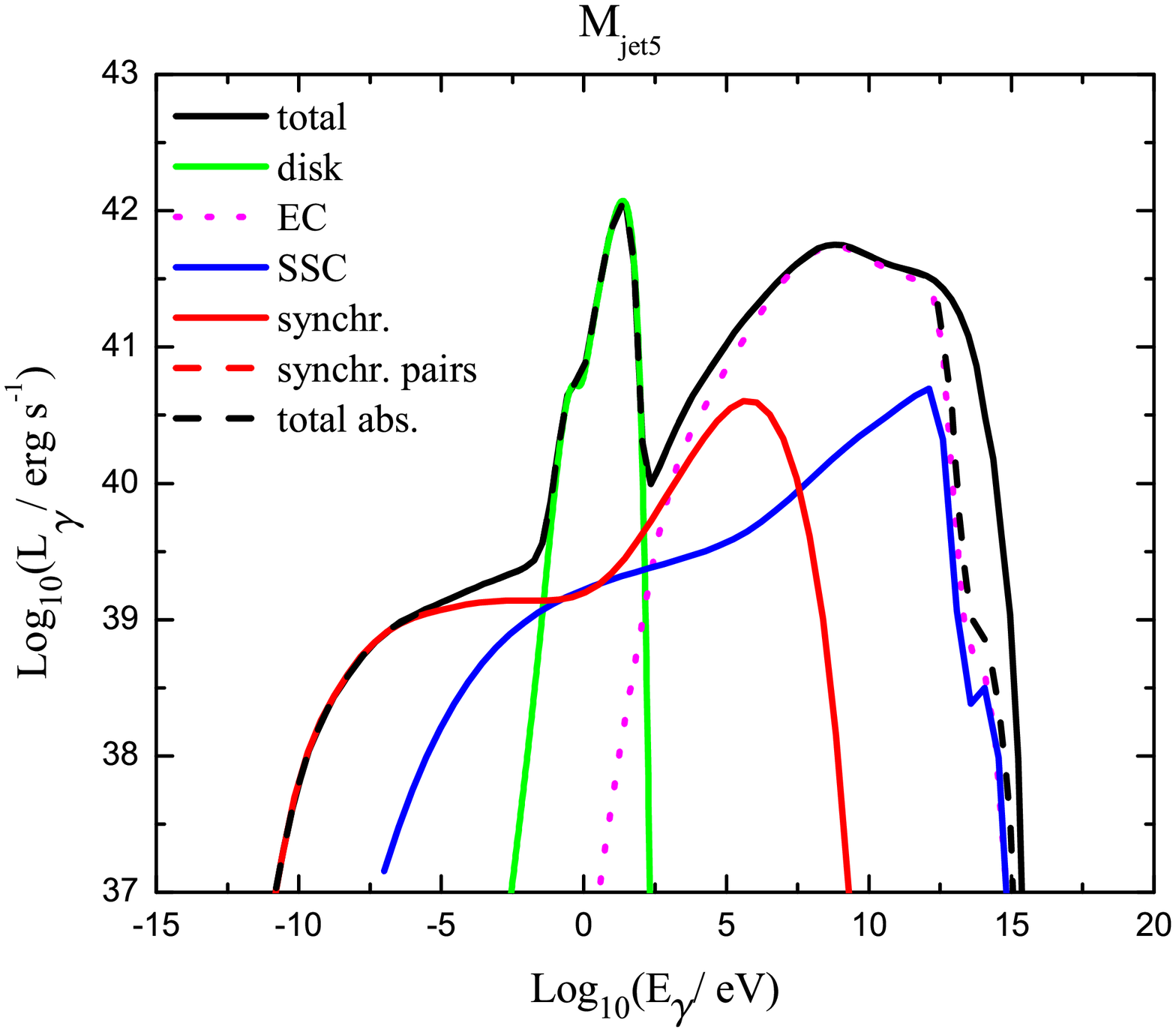}
\includegraphics[width = 0.48\textwidth, keepaspectratio]{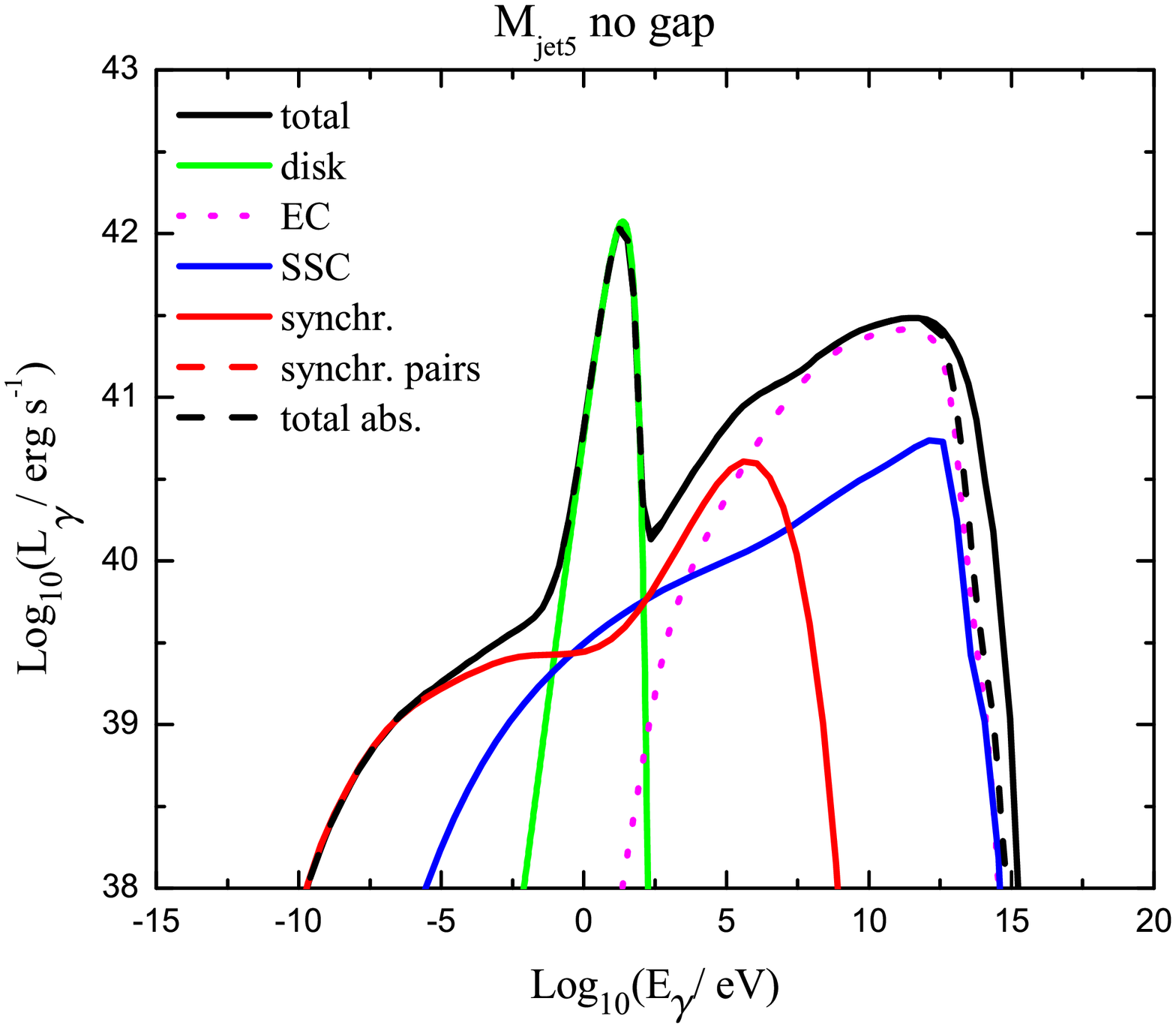}
\caption{Left: spectral energy distribution of the jet corrected for absorption for the set of parameters $M_{\rm jet5}$. Right: the same but for the case of a relativistic thin disk without a gap around a black hole of the same mass, spin, and accretion rate as the primary.  }
\label{fig:sed_jet_M5}
\end{figure*}

\begin{figure*}[htbp]
\center
\includegraphics[width = 0.48\textwidth, keepaspectratio]{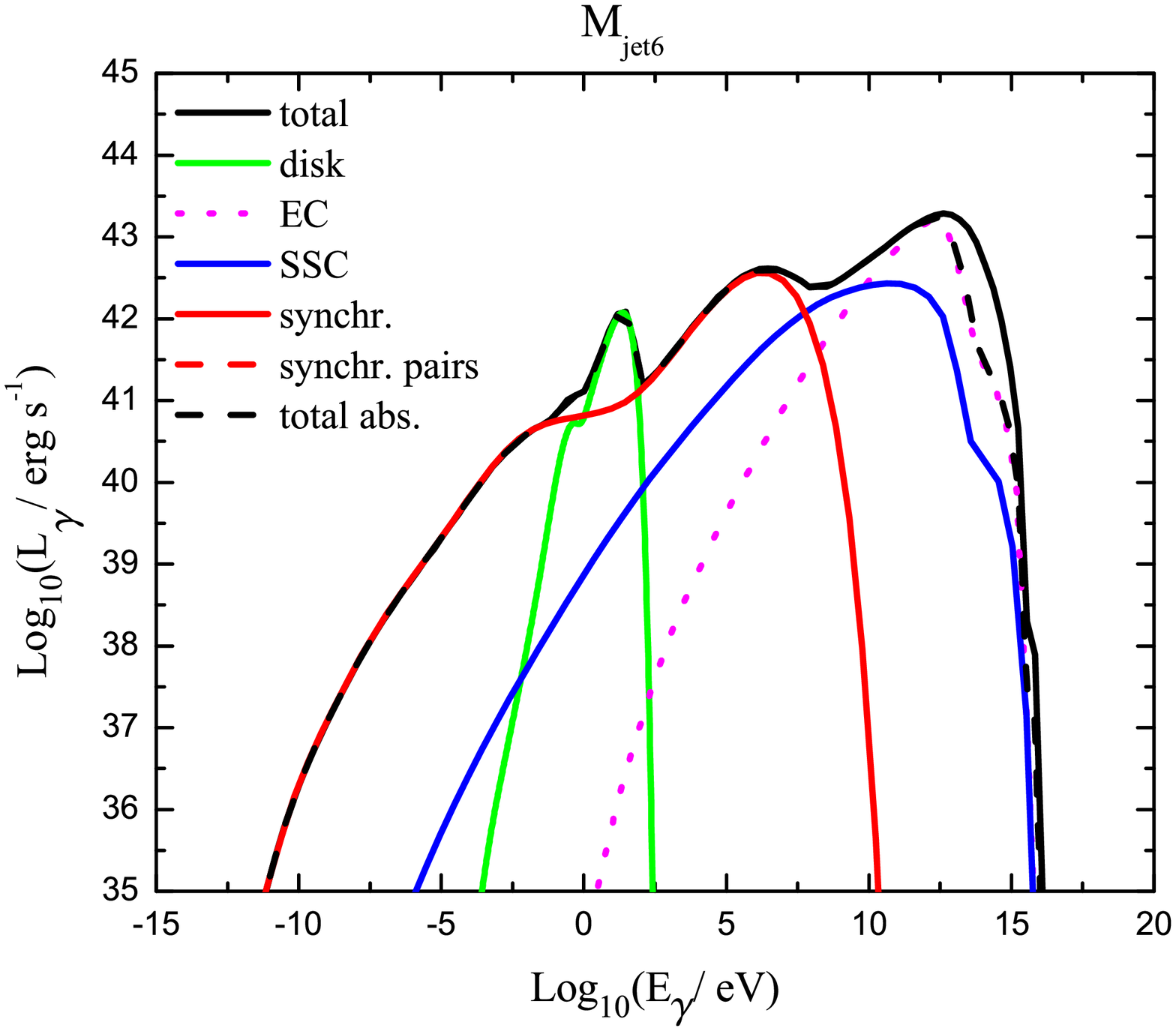}
\includegraphics[width = 0.48\textwidth, keepaspectratio]{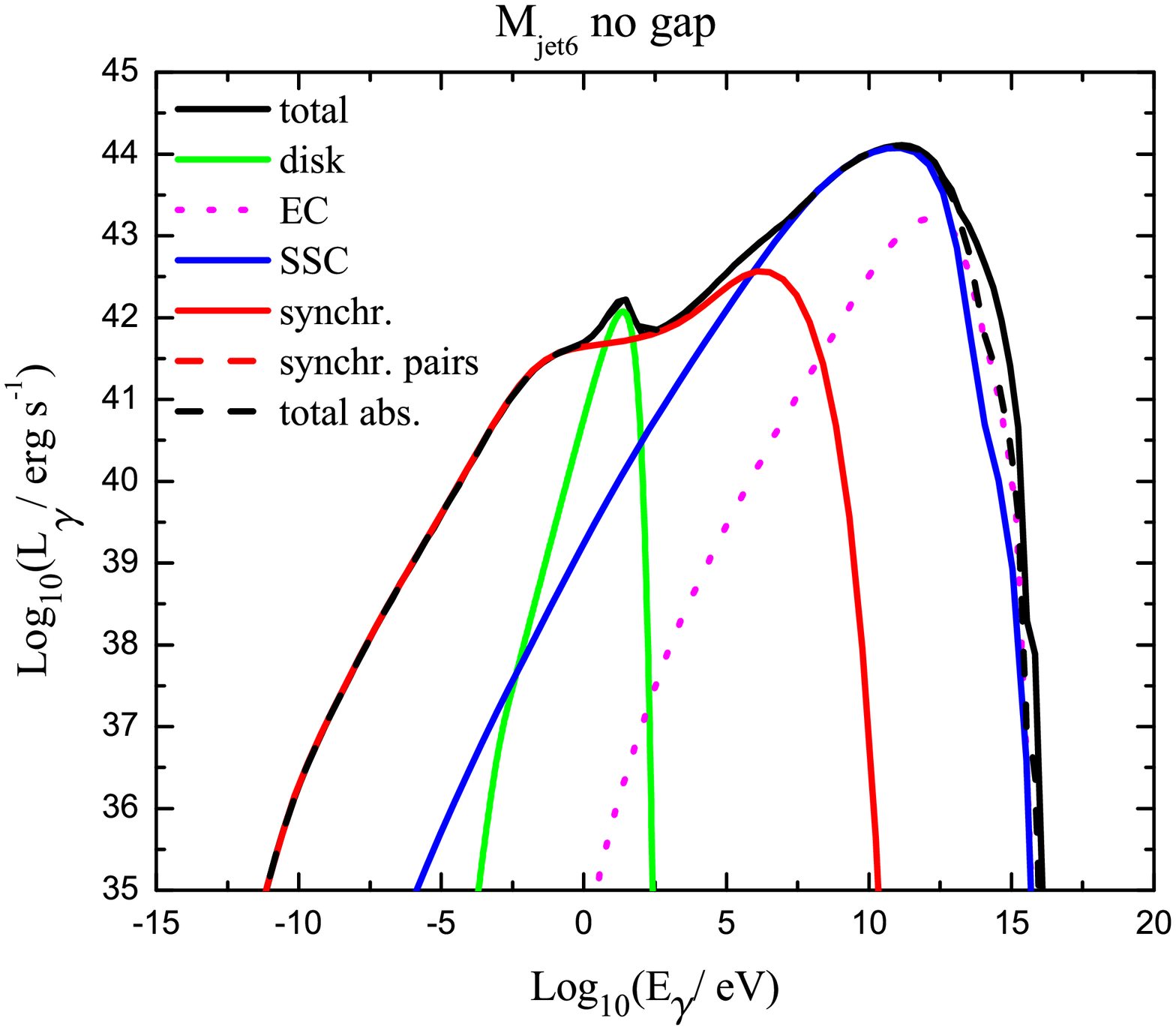}
\caption{Left: spectral energy distribution of the jet corrected for absorption for the set of parameters $M_{\rm jet6}$. Right: the same but for the case of a relativistic thin disk without a gap around a black hole of the same mass, spin, and accretion rate as the primary.  }
\label{fig:sed_jet_M6}
\end{figure*}

We have calculated the jet emission for a large number of models  characterised by different sets of parameters in order to explore the impact of the existence of the gap and modified external thermal radiation on the SEDs of different systems. In Figs. \ref{fig:sed_jet_M1}-\ref{fig:sed_jet_M6} we show a sample of cases that illustrate the broadband spectral energy distributions of the jet radiation of binary black hole systems with a gap in their disks. The set of parameters for the jet models are listed in Table \ref{T2}. Models are labelled from $M_{\rm jet1}$ to  $M_{\rm jet6}$. The corresponding disk models are $M_{\rm disk2}$ for  $M_{\rm jet1}$, and $M_{\rm disk1}$ for the remaining cases. For comparison, the SEDs obtained for the same jets parameters but for the case of a disk without a gap are also plotted along the SEDs shown in Figs.  \ref{fig:sed_jet_M1}-\ref{fig:sed_jet_M6}.

In all jet models, the base of the jet is close to the black hole (50 gravitational radii) and the particle acceleration zone extends from the near base region up to several thousands of gravitational radii. The jet bulk Lorentz factor is moderate in the first three models (about 20) and slower in the others (from 5 to 10, see Table \ref{T2}); for a general reference about this see \citet{MOJAVE2009}. The original electron injection is canonical with a slope of $2.2$ in all models, except in model  $M_{\rm jet6}$, which has a hard spectral index of $1.5$. Viewing angles are small (from 5 to 1 degree) so we are dealing with blazar-type objects. The jet power is in the range from $\sim10^{41}- \;\sim 10^{44}$ erg s$^{-1}$. All results presented in the figures have been corrected by internal absorption of gamma rays and are shown as seen in the observer's frame. 

At low energies, all SEDs are dominated by synchrotron radiation. At optical-UV energies, since the jets are not exceedingly powerful, the blue bump with the characteristic shape imprinted by the gap is  discernible in most cases. At high energies, the emission is the result of the addition of EC and SSC. We  selected models dominated by EC. All models present a softening in the gamma-ray spectrum around 10 GeV when compared with the corresponding models without a gap. The suppression of the very high-energy radiation by photon annihilation also occurs at lower energies in the models with a gap than in those with standard disks. This is the effect of the excess of soft photons coming from the outer border of the gap in the perturbed disks. 

The model with greatest differences with respect to unperturbed disks is model $M_{\rm jet6}$. This is the case of the hardest injection we have considered. The synchrotron peaks at X-rays, as in X-ray selected BL Lac objects. Three peaks can be distinguished in the SED: the disk, the synchrotron peak, and the EC peak. On the contrary, in the corresponding case without gap, a hard spectrum extends all the way from UV to TeV energies. 

We also notice that models with primary of the order of \mbox{$10^7$ $M_{\odot}$} display more peculiar high-energy features than models with more massive black holes. In these disks the near exterior region is located farther away from the black hole; what causes that the excess in the emission of the perturbed disk is shifted towards lower photon energies, as seen in Figure \ref{fig:sed_disk}. In these models, then, there is an excess of target disk photons in the appropriate energy band to produce gamma rays when they are up-scattered by the relativistic electrons in the jets.

The secondary to primary black hole ratio for the models with the strongest specific high-energy features, which can be used to trace binarity, is $\sim 0.05$. So the secondary is an intermediate black hole captured by the primary. Old starbursts whose central black hole has been reactivated by a fresh inflow of gas, hence, might be propitious progenitors of these kinds of binary systems.   

Other regions of the parameter space allowed by the ``overflowing'' regime remain to be explored, but we expect the effects of the existence of the gap are maximised for values of $M_\bullet$ in the range analysed here. For less massive primaries, the gap would be wider but located too far out in the coldest  regions of the disk. For more massive primaries, on the other hand, the secondary would be nearer but the gap would be too narrow to induce any noticeable feature in the SED.

\begin{table*}
\caption{Values of the relevant model parameters.}
\centering
\begin{tabular}{l | c  | c }
\hline  
Parameter, symbol [units]  & $M_{\rm disk1} $ & $M_{\rm disk2}$  \\ [0.1cm]
\hline	
Mass of the primary, $M_{\bullet}$  [$M_\odot$]          & $10^7$                   & $10^8$              \\[0.1cm]
Mass accretion rate, $\dot{M}$  [$\dot{M}_{\rm Edd}$] & $0.12$  & $0.12$ \\[0.1cm]
Mass ratio, $q$ & $5.5\:10^{-3}$  &  $2.4\:10^{-3}$   \\[0.1cm]
Orbital separation, $r_{\rm s}$ [$r_{\rm g}$] & $1.8\:10^{4}$  &  $1.5\:10^{3}$   \\[0.1cm]
Hill radius (gap half width), $r_{\rm H}$ [$r_{\rm g}$] & $2.2\:10^{3}$  &  $1.4\:10^{2}$   \\[0.1cm] 
Disk inner radius (radius of the ISCO), $r_{\rm in}$ [$r_{\rm g}$] & $1.45$  &  $1.45$   \\[0.1cm]
Transition radius inner far/near interior regions, $r_{\rm ni}$ [$r_{\rm g}$] & $1.2\:10^{4}$  &  $1.1\:10^{3}$   \\[0.1cm] 
Transition radius near exterior/middle regions, $r_{\rm ne}$ [$r_{\rm g}$] & $5.6\:10^{4}$ &   $3.9\:10^{3}$\\[0.1cm] 
\hline
\end{tabular}
\label{tab:general_parameters}
\end{table*}

\begin{table*}
\caption{Values of the parameters for the six disk+jet models.}
\centering
\begin{tabular}{l | c  | c | c}
\hline  
Parameter, symbol [units]  &  $M_{\rm jet1}$ &  $M_{\rm jet2}$ & $M_{\rm jet3}$ \\ [0.1cm]
\hline	
Disk model & $M_{\rm disk2}$ & $M_{\rm disk1}$ & $M_{\rm disk1}$ \\ [0.1cm]
Base of the jet, $z_{0}$ [$r_{\rm g}$] & $50$  &  $50$  &  $50$  \\[0.1cm]
Base of the acceleration region, $z_{\rm acc}$ [$r_{\rm g}$] & $70$ & $70$ & $70$  \\[0.1cm]
End of the acceleration region, $z_{\rm max}$ [$r_{\rm g}$]  & $7000$ & $3000$  &  $3000$\\[0.1cm]    
Termination of the jet, $z_{\rm end}$ [$r_{\rm g}$] & $7\:10^{5}$ & $7\:10^{6}$  & $7\:10^{6}$ \\[0.1cm]
Jet inclination angle, $\theta_{\rm jet}$ [deg] & $4$ &  $0$ & $4$ \\[0.1cm]
Jet viewing angle, $\theta_{\rm obs}$ [deg] & $1$ &  $5$ & $1$\\[0.1cm]
Jet bulk Lorentz factor, $\Gamma_{\rm jet}$ & $20$  & $20$ &  $20$  \\[0.1cm]
Magnetic field at $z_{\rm acc}$, $B(z_{\rm acc})$ [G] &  $0.7$ & $2.1$  & $2.1$ \\[0.1cm]
Jet power, $L_{\rm jet}$ [${\rm erg}\:{\rm s}^{-1}$]  & $6.5\:10^{43}$  &  $6.5\:10^{42}$  & $6.5\:10^{42}$ \\[0.1cm]
Power relativistic electrons, $L_{\rm rel}$ [${\rm erg}\:{\rm s}^{-1}$]  & $3.2\:10^{42}$  & $3.2\:10^{41}$ & $3.2\:10^{41}$\\[0.1cm]
Minimum energy relativistic electrons, $E_{\rm min}$ [$m_e c^2$]  & $10$  & $10$ &  $10$ \\[0.1cm]
Injection spectral index, $p$ & $2.2$  & $2.2$ &  $2.2$  \\[0.1cm]
Acceleration efficiency, $\eta$ & $0.1$  & $0.1$ &  $0.1$\\[0.1cm]
\hline\hline
Parameter, symbol [units]  &  $M_{\rm jet4}$ &  $M_{\rm jet5}$ & $M_{\rm jet6}$\\ [0.1cm]
\hline	
Disk model &  $M_{\rm disk1}$ & $M_{\rm disk1}$ & $M_{\rm disk1}$\\ [0.1cm]
Base of the jet, $z_{0}$ [$r_{\rm g}$] & $50$  &  $50$  &  $50$ \\[0.1cm]
Base of the acceleration region, $z_{\rm acc}$ [$r_{\rm g}$] & $70$ & $70$ & $70$ \\[0.1cm]
End of the acceleration region, $z_{\rm max}$ [$r_{\rm g}$] &  $3000$  & $3000$ & $3000$\\[0.1cm]    
Termination of the jet, $z_{\rm end}$ [$r_{\rm g}$] & $7\:10^{6}$ &  $7\:10^{6}$ & $7\:10^{6}$ \\[0.1cm]
Jet inclination angle, $\theta_{\rm jet}$ [deg] & $0$ & $0$ & $0$ \\[0.1cm]
Jet viewing angle, $\theta_{\rm obs}$ [deg] & $5$ & $5$ & $5$\\[0.1cm]
Jet bulk Lorentz factor, $\Gamma_{\rm jet}$ & $10$ & $5$ & $10$\\[0.1cm]
Magnetic field at $z_{\rm acc}$, $B(z_{\rm acc})$ [G] &  $2.1$ &   $2.1$ & $2.1$ \\[0.1cm]
Jet power, $L_{\rm jet}$ [${\rm erg}\:{\rm s}^{-1}$]  & $6.5\:10^{42}$ & $6.5\:10^{42}$ & $6.5\:10^{42}$\\[0.1cm]
Power relativistic electrons, $L_{\rm rel}$ [${\rm erg}\:{\rm s}^{-1}$]  & $3.2\:10^{41}$  & $3.2\:10^{41}$ &  $3.2\:10^{41}$\\[0.1cm]
Minimum energy relativistic electrons, $E_{\rm min}$ [$m_e c^2$]  & $10$  & $10$ &  $10$\\[0.1cm]
Injection spectral index, $p$ & $2.2$  & $2.2$ &  $1.5$  \\[0.1cm]
Acceleration efficiency, $\eta$ & $0.1$  & $0.1$ &  $0.1$\\[0.1cm]
\hline
\end{tabular}
\label{T2}
\end{table*}

\section{Discussion}
\label{sect:discussion}

\subsection{Robustness and detectability}

The kind of model studied in this paper is intentionally tuned to maximise the observational features of SMBHBs in AGN. Although several conditions should be fulfilled by a given system to be detected through its high-energy emission, most of these conditions are not unlikely to be met by several objects in a large sample. 

Specifically, we request the following conditions to be fulfilled. Firstly, the binary orbit of the black hole system must be co-planar. This is expected to be the predominant case among supermassive binaries because the process of resonant and hydrodynamic damping of inclination of a satellite (secondary) object interacting with a circumprimary disk (e.g. \citealt{Artymowicz1998}). These effects are well known from planetary formation.

 Then, the mass ratio between the secondary and primary black holes must be rather small (in the range $0.1-10^{-3}$) to secure a large gap. This case does not seem to be unusual since primary black holes with masses in the range \mbox{$10^7 - 10^8$ $M_{\odot}$} are thought to power most AGN and intermediate black holes should be common in satellite galaxies. Mergers at high redshifts were also a common occurrence (\citealt{WilsonColbert1995}). 
 
 In our models, the separation between the black holes is of several thousands of gravitational radii of the primary. This ensures that the excess of radiation in the perturbed disk is located between the IR and optical bands, which is a suitable target for EC scattering in the jet. Very close binaries will be short-lived because of the strong gravitational wave losses. The medium-separation systems are then expected to be the most numerous among SMBHBs, so this requirement should be met for most binaries.  
 
 Another assumption of our model is that the secondary is able to open a gap in the disk. This is met essentially by all systems that also meet the conditions mentioned above, as shown by the simulations performed by \cite{Kocsis-etal2012a,Kocsis-etal2012b}; see also Sect. \ref{sec:disk_model}. 
 
 We demand that the eccentricity of the SMBHB is low, so the circular orbit approximation is valid. This requirement is expected to be met by most systems because of the rapid circularisation of the orbits produced by the viscous interactions between the secondary and gas in the disk. This effect is well known from planetary disks. 
 
 In order to have jets and, hence, gamma-ray emission, accretion onto the primary must proceed in the overflowing regime (Type 1.5 migration). The numerical simulations of \citet{Kocsis-etal2012b,Kocsis-etal2012a} show that this case is standard for  the range of mass ratios considered in our calculations. The fact that in the models the AGN launches a relativistic jet that points nearly in the direction of the observer is satisfied essentially by all non-nearby AGN seen at gamma rays; otherwise, the sources would not be detectable. 
 
 A crucial requirement of our models is that particles are accelerated in a jet not too far from the central source. Otherwise, EC scattering would not be effective. This does not seem to be a problem, since the vast majority of blazars have a high incidence of fast variability, which indicates that the radiation is being produced in a very compact region close to the central black hole (e.g. \citealt{Romeroetal1997, Romeroetal1999, Romeroetal2002}). 
 
 Finally, the EC process is the dominant contribution at gamma rays. This dominance depends basically on the magnetic field in the jet: where the field is high, electrons cool mainly by synchrotron and SSC mechanisms. The detectability of binarity through the features described in this paper crucially depends, then, on having a rather modest magnetic field strength $B \lesssim 1$ G in the gamma-ray emitting region. Since this region extends up to distances of several thousands of gravitational radii and the field must be below equipartition in order for the fluid to remain compressible (otherwise shocks would not develop), we can expect such values not to be exceptional. 

 So, altogether, how robust is our model and what are the real chances of detecting a SMBHB through gamma-ray observations? A quantitative estimate is not possible, since some of the conditions numbered above remain essentially qualitative because of lack of a knowledge about galaxy evolution. However, with several thousands of AGN detected so far by the {\it Fermi} gamma-ray telescope and Cherenkov arrays, the actual chance of having already observed one SMBHB with its SED dominated by EC radiation should be close to one. Of course, to identify such a source in the sample is a very different issue, and the SED calculations presented in this paper are aimed at providing templates for such a search.

\subsection{Dependence on changes in the parameter space}

How sensitive is our model to changes in the different parameters? In particular the magnetic field and the location of the acceleration region? In order to answer this question we have computed a number of SEDs systematically changing these parameters. The following general trends were observed.  

\begin{figure*}[htbp]
\centering
\includegraphics[width = 0.49\textwidth, keepaspectratio]{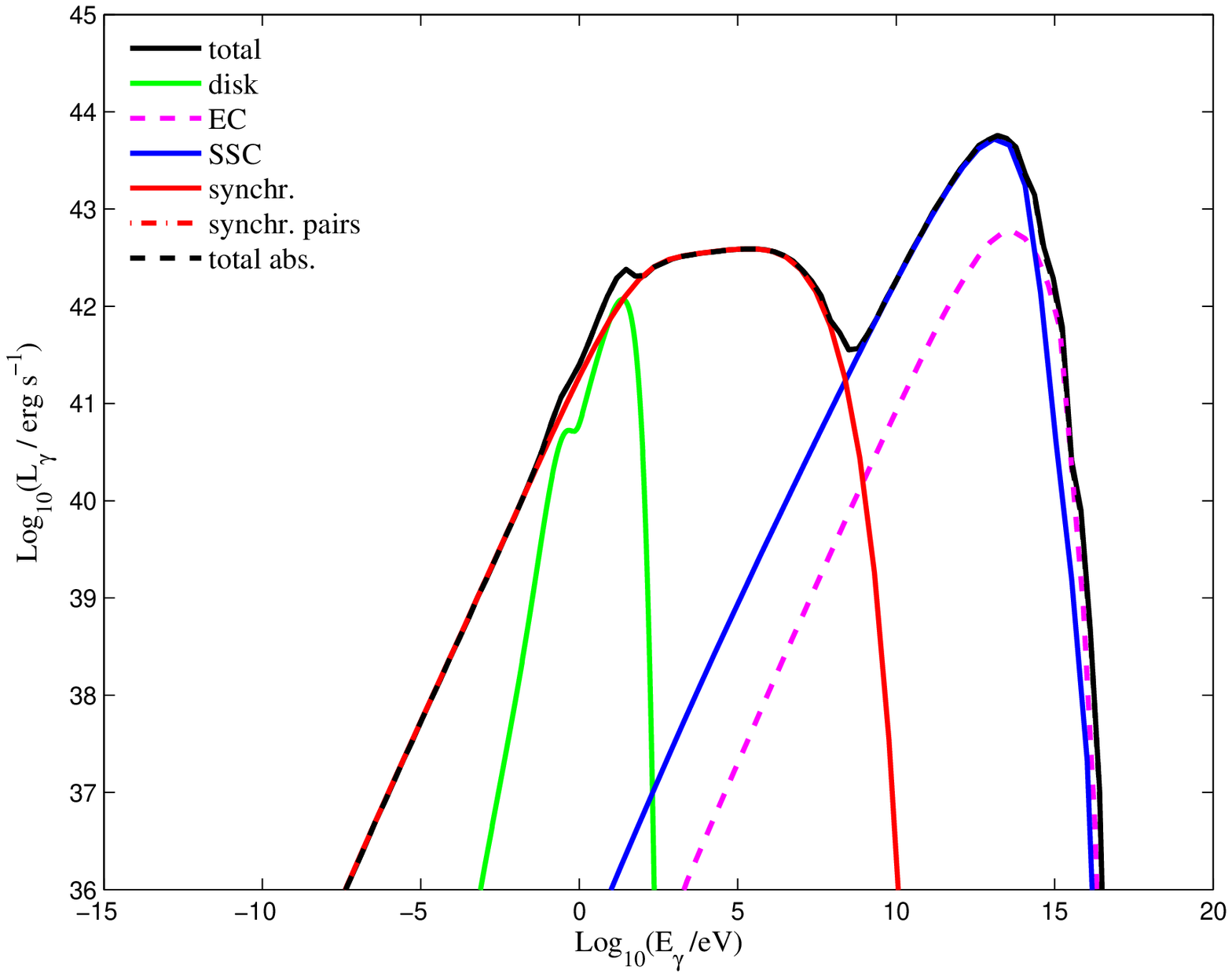}
\includegraphics[width = 0.49\textwidth, keepaspectratio]{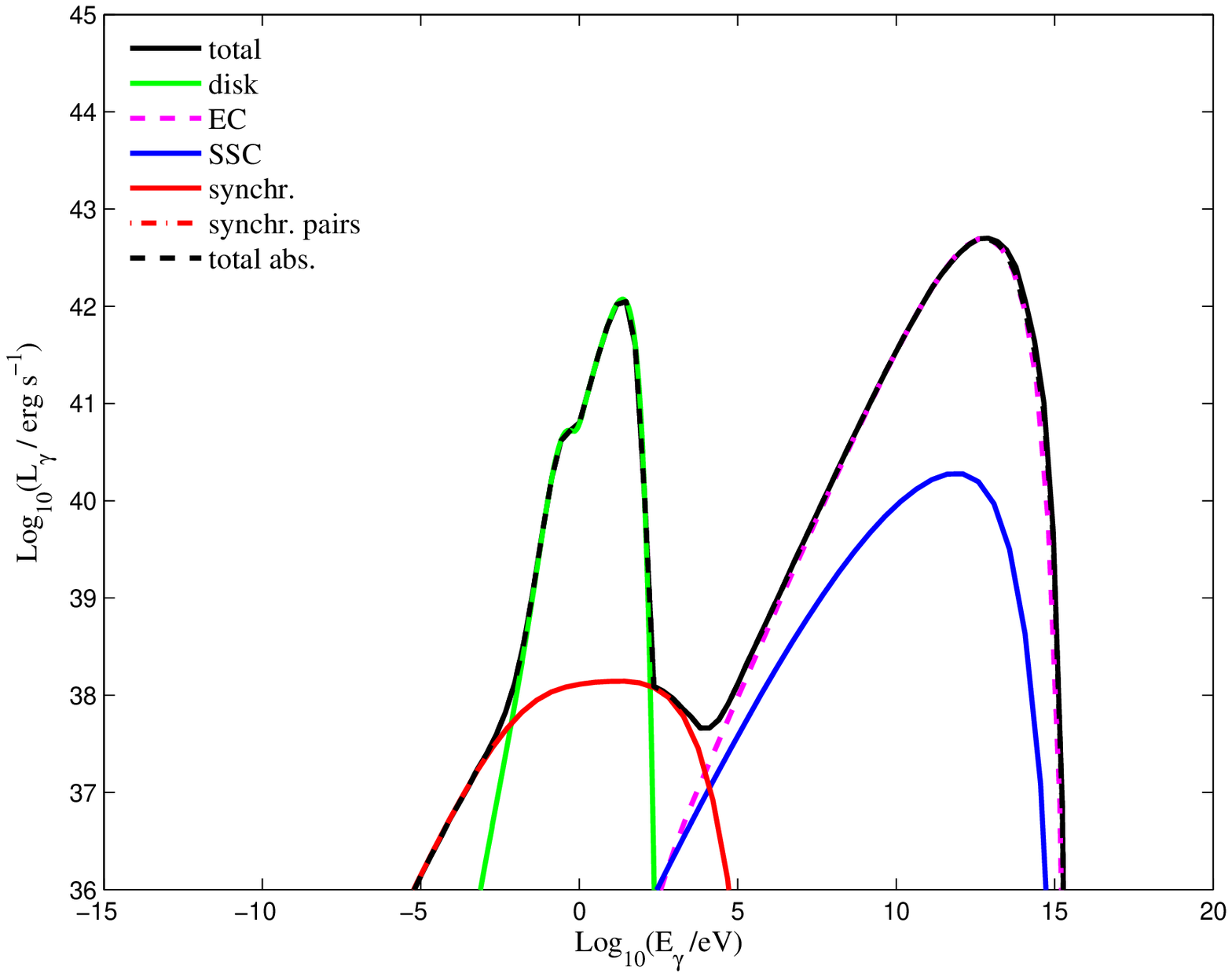}\\
\includegraphics[width = 0.49\textwidth, keepaspectratio]{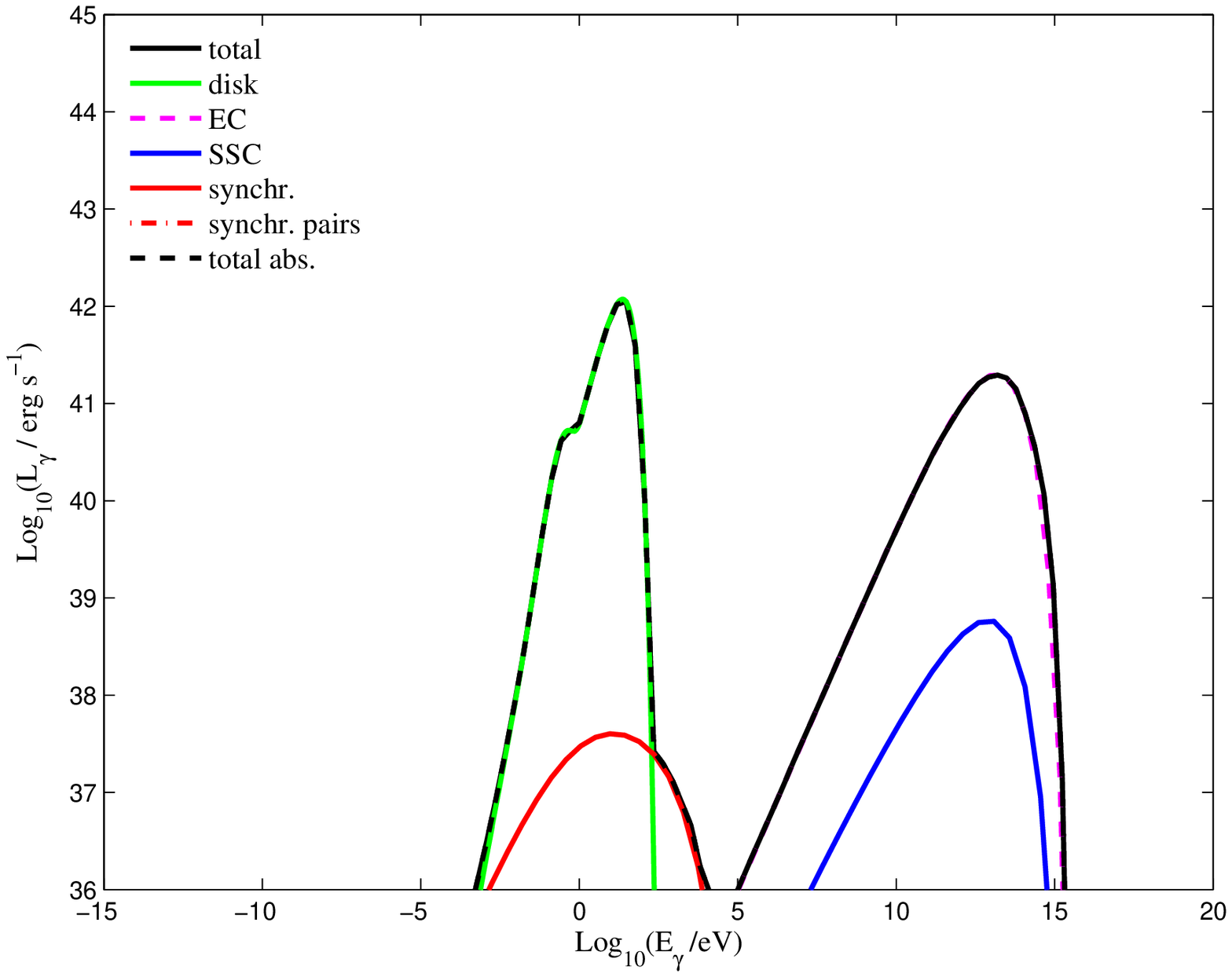}
\includegraphics[width = 0.49\textwidth, keepaspectratio]{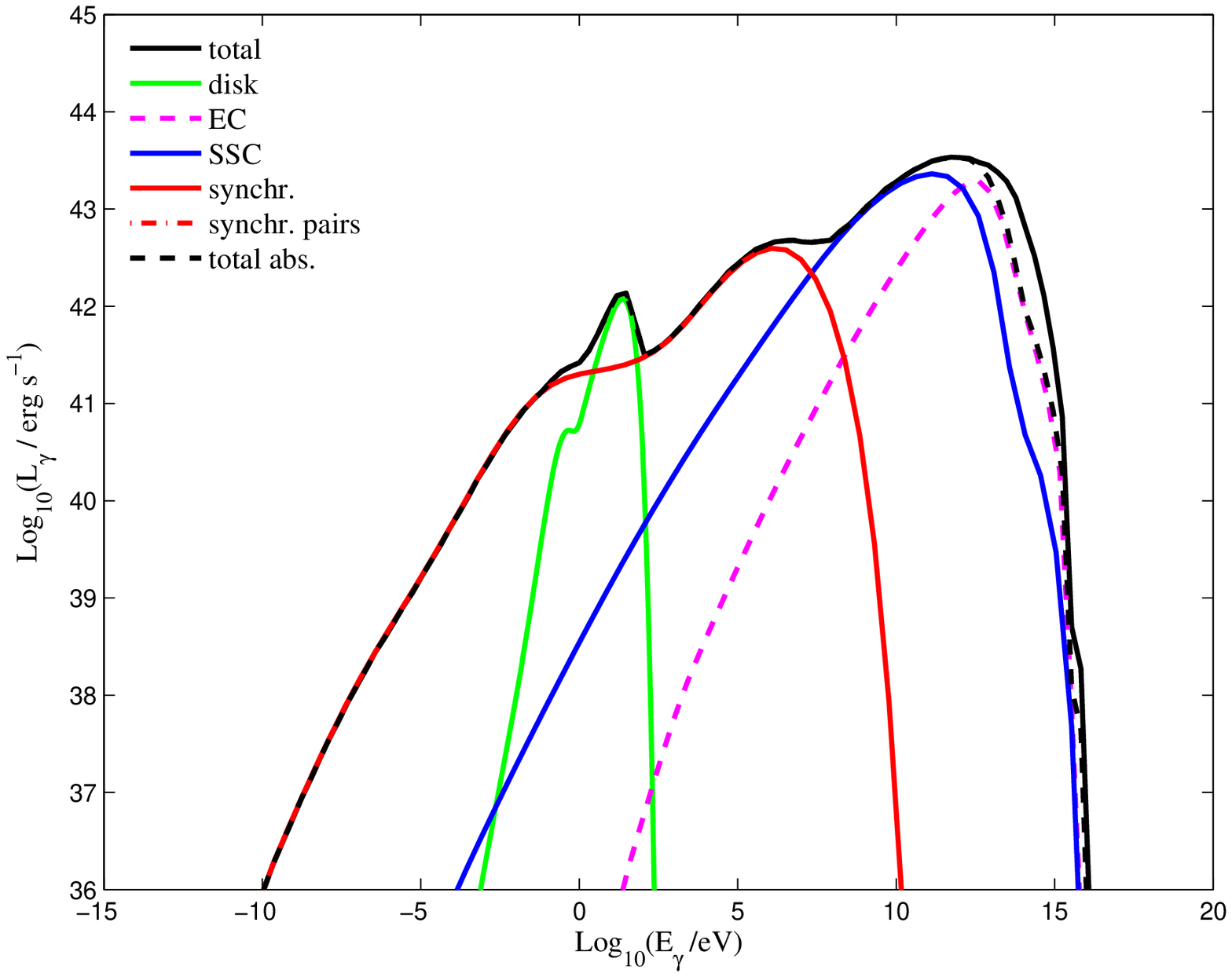}
\caption{Changes in the SED with changes in some of the basic parameters of the basic model. In all cases, $z_0=50\;r_{\rm g}=z_{\rm acc}$, $p=1.5$, and $\Gamma_{\rm jet}=10$. The upper panels, from left to right, show models with $z_{\rm max}=10^3 z_0$ but differ in the value of the magnetic field at the base of the acceleration region: $B(z_{\rm acc}) = 1.8$ G and $B(z_{\rm acc}) =0.06$ G, respectively. The models in the lower panels differ in the extension of the acceleration region: $z_{\rm max} = 10^4 z_0$ and $10^2 z_0$, respectively. In both cases $B(z_{\rm acc})=0.06$ G.
 }
\label{fig:seds_change_parameters}
\end{figure*}

As expected, a high value of the magnetic field results in a strong synchrotron component; the associated high-energy SSC peak in the SED is also enhanced (see Fig. \ref{fig:seds_change_parameters}, upper left panel).
Models that have similar conditions but have a field lower by two orders of magnitude have the synchrotron peak suppressed and the high-energy radiation dominated by the EC component. We notice that these models have compact acceleration regions located close to the base of the jet, and hence, there are plenty of photons coming from the disk for IC interactions  (see Fig. \ref{fig:seds_change_parameters}, upper right panel). If the acceleration region extends to distances larger by an order of magnitude, two effects are observed: the synchrotron peak almost disappears and the EC peak at high energies is also strongly diminished. The reason is that the same budget of energy in relativistic particles is distributed along a much broader region, the farthest parts of which are at such a distance from the disk the EC scattering becomes ineffective (see Fig. \ref{fig:seds_change_parameters}, lower left panel). The models with the strongest high-energy features are those where the acceleration region is very compact and located at the base of the jet. In these models, the whole energy budget of relativistic particles is injected very close to the source of external photons and, at the same time, synchrotron losses tend to be high because of relatively high values of the magnetic field at the base of the jet. Such a combination of parameters results in a broad two-peaked SED at high energies, as shown in Fig. \ref{fig:seds_change_parameters}, lower right panel.

The existence of an acceleration region close to the base of the jets seems to be favoured in most blazars by the observation of fast variability, which is indicative of a very compact source. At the same time, very strong magnetic fields are implausible because if the jet flow is magnetically dominated, shocks are not produced since the gas is mechanically incompressible. In all models we computed in this work the magnetic field is below equipartition with the gas.

\section{Closing remarks}

We have studied the radiative signatures of SMBHBs. We have focused on a particular kind of accreting binaries: those where the primary is more massive and the secondary can open a gap on the accretion disk of the system. We have shown that when a relativistic jet is present, external Compton interactions between photons from the perturbed disk and relativistic electrons in the base of the jet, can produce a unique signature in the spectral energy distribution at gamma-ray energies, where all the emission is of non-thermal origin. Admittedly, not all binary cores in AGN will display these features. However, the mere identification of a single system might provide a unique natural laboratory to study black hole binaries and the related accretion physics through high-energy observations. 

The SEDs  obtained in this work differ from the typical spectra with two bumps well fit by power laws predicted by one-zone models of AGN jets. In one-zone models, the electron distribution $N(E,z)$ is usually a single power law in energy for all $z$, whether fixed beforehand or calculated from simplified versions of Eq. \ref{eq:transport_equation}, which assume a dominant cooling mechanism. In our model the emission region is extended, inhomogeneous, and the dominant process of energy loss of the electrons varies along the jet. The shape of the function $N(E,z)$ is thus very complex and this directly shows in the SEDs. We also notice that a value of the magnetic field in the acceleration region of \mbox{ $\sim 1$ G  or less} is required to avoid the synchrotron component of the SED masking the EC component. Such fields are usually estimated for the inner sub-parsec jet (e.g. \citealt{MarscherGear1985, DermerSchlickeiser1993, Romero-etal1995}). 

Differences in the low-energy cutoff of the EC emission do not produce noticeable effects. Nevertheless, optical and IR observations of specific objects might help to find direct signatures of the disks. According to our results, the characteristics of the SED at very high energies (alongside the features expected in the emission of the accretion disk in the optical) could provide a good criterion for the identification of SMBHB candidates. In this energy range, future observations with  Cherenkov telescopes, such as the Cherenkov Telescope Array (CTA; \citealt{Actis-etal2011}), will be an important tool to investigate the effects predicted here.

Once a set of good SMBHB candidates has been identified, future space-borne gravitational wave detectors can attempt to find the metric waves produced by these systems.   

\begin{acknowledgements}
Early stages of this research were supported by grants PIP 0078/2010 (CONICET) and PICT 2012-00878 Pr\'estamo BID (ANPCyT). Black hole astrophysics with G. E. Romero is also supported through grant AYA 2013-47447-C3-1-P from MEyC (Spain). G. E. Romero {\it should} have been assisted by a so-far unpaid new grant from CONICET.  
\end{acknowledgements}

\bibliographystyle{aa}	
\bibliography{myrefs}		

\end{document}